\documentclass[11pt]{article}
\usepackage{latexsym,amssymb,amstext,amsmath}
\usepackage{hyperref}
\usepackage{graphicx}
\usepackage{amsmath} 
\usepackage{color}
\allowdisplaybreaks

\begin{document}
%
\begin{titlepage}
\begin{flushright} \small
Nikhef-2019-024 
\end{flushright}
\bigskip

\begin{center}
 {\LARGE\bfseries Exact results for an  STU-model}
\\[10mm]

\textbf{G.L.~Cardoso$^{a}$, B.~de~Wit$^{b,c}$ and S.~Mahapatra$^{d}$}

\vskip 6mm
$^a${\em  Center for Mathematical Analysis, Geometry and Dynamical
  Systems,\\ 
  Department of Mathematics, 
  Instituto Superior T\'ecnico,\\ Universidade de Lisboa,
  Av. Rovisco Pais, 1049-001 Lisboa, Portugal}\\[.2ex]
$^b${\em Nikhef, Science Park 105, 1098 XG Amsterdam, The
  Netherlands}\\[.2ex] 
$^c${\em Institute for Theoretical Physics, Utrecht University,\\
   Princetonplein 5, 3584 CC Utrecht, The Netherlands}\\[.2ex]
$^d${\em Physics Department, Utkal University, 
Bhubaneswar 751 004, India}\\[1ex]

{\tt gcardoso@math.tecnico.ulisboa.pt}\;,\;\,{\tt  B.deWit@uu.nl}\;,\;\,{\tt
  swapna@iopb.res.in}
\end{center}

\vskip .2in
\begin{center} {\bf ABSTRACT } \end{center}
\begin{quotation}\noindent 
  The duality symmetries of the STU-model of Sen and Vafa are very
  restrictive. This is utilized to determine the holomorphic function
  that encodes its two-derivative Wilsonian effective action and its
  couplings to the square of the Weyl tensor to fifth order in
  perturbation theory.  At fifth order some ambiguities remain which
  are expected to resolve themselves when proceeding to the next
  order.  Subsequently, a corresponding topological string partition
  function is studied in an expansion in terms of independent
  invariants of $S$, $T$ and $U$, with coefficient functions that
  depend on an effective duality invariant coupling constant $u$,
  which is defined on a Riemann surface $\mathbb{C}$. The coefficient
  function of the invariant that is independent of $S$, $T$ and $U$ is
  determined to all orders by resummation. The other functions can be
  solved as well, either algebraically or by solving differential
  equations whose solutions have ambiguities associated with
  integration constants.  This determination of the topological string
  partition function, while interesting in its own right, reveals new
  qualitative features in the result for the Wilsonian action, which
  would be difficult to appreciate otherwise. It is demonstrated how
  eventually the various ambiguities are eliminated by comparing the
  results for the effective action and the topological string. While
  we only demonstrate this for the leading terms, we conjecture that
  this will hold in general for this model.

\end{quotation}
\vfill
\end{titlepage}

\section{Introduction}
\label{sec:introduction}
\setcounter{equation}{0}

For a general $N=2$ string compactification it is difficult to obtain
exact expressions for the part of the Wilsonian effective action that
describes the gravitational interactions with the vector multiplets,
as well as for the topological string partition function. Often one
has to make use of partial results obtained by concentrating on the
neighbourhood of special points in the string moduli space, or from
integrating the holomorphic anomaly equation for low genus. In this
paper we want to investigate whether there exists a model for which
the Wilsonian action and the topological string partition function can
in principle be derived from their duality symmetries. The model we
have in mind was discovered by Sen and Vafa when constructing dual
pairs of type-II string compactifications in four space-time
dimensions with $N=2$ supersymmetry \cite{Sen:1995ff}.  These pairs
were obtained by appropriate $\mathbb{Z}_2 \times \mathbb{Z}_2$
orbifold constructions based on toroidally compactified type-II string
theory. One such dual pair (referred to as N=2 Example D) is described
by an $N=2$ supergravity model with three vector multiplets and four
hypermultiplets.

This model, which we will call the STU-model in the following, is the
subject of study in this paper. It has a type-II description based on
a Calabi-Yau three-fold with vanishing Euler number.  In this
description, the dilaton belongs to a hypermultiplet, and therefore
the vector moduli space does not receive quantum corrections. The
exact vector moduli space is based on an
$\big[\mathrm{SL(2)}/\mathrm{SO(2)} \big]^3$ coset space with each
factor modded out by the action of the integer-valued subgroup
$\Gamma_0(2)\subset \mathrm{SL}(2;\mathbb{Z})$, defined by restricting
its integer-valued matrix elements $a,b,c,d$ with $ad-bc=1$, by
$a, d\in 2\,\mathbb{Z}+1$, $c\in 2\,\mathbb{Z}$ and $b\in
\mathbb{Z}$. The quantum moduli space is therefore equal to
$\big[\Gamma_0(2)\backslash\mathrm{SL}(2)/\mathrm{SO}(2)\big]^3$, and
the vector multiplet sector is invariant under the product of an S-, a
T- and a U-duality group,
$\Gamma_0(2)_\mathrm{S}\times \Gamma_0(2)_\mathrm{T}\times
\Gamma_0(2)_\mathrm{U}$. The role of the group $\Gamma_0(2)$ followed
from the determination of the one-loop gravitational coupling that was
computed explicitly by working at an orbifold point of the Calabi-Yau
three-fold \cite{Gregori:1999ns}. This coupling is invariant under
$[\Gamma_0(2)]^3$ and involves the logarithm of a modular form that
will play an important role in this paper. In addition, the model is
also invariant under arbitrary permutations of the vector moduli
$S, T$ and $U$. We will refer to this symmetry as {\it triality}. We
stress that this is a different STU-model than the one studied, for
instance, in \cite{Grimm:2007tm}, which is a reduction of the
FHSV-model and whose duality group equals
$\mathrm{SL}(2, \mathbb{Z})_\mathrm{S}\times \Gamma
(2)_\mathrm{T}\times\Gamma (2)_\mathrm{U}$.
 
 In this paper we will assume that the STU-duality symmetries
 (triality and $[\Gamma_0(2)]^3$ symmetry) remain valid at any order
 in perturbation theory, and we will use these symmetries to determine
 the holomorphic function \cite{deWit:1984pk} which encodes the part
 of the Wilsonian effective action that includes interactions
 proportional to the square of the Weyl tensor to fifth order in its
 gravitational coupling.  Here it is important to stress that
 generically such functions do not transform as a function under
 duality transformations, and they are thus not straightforwardly
 invariant under the duality transformations of the moduli. Rather,
 the higher-order terms that one wishes to include should be such that
 the action of the electric-magnetic dualities on the field strengths
 and their duals will remain the same
 \cite{Cecotti:1988qn,Cardoso:2008fr,Cardoso:2010gc,Cardoso:2012nh}. In
 the case at hand the holomorphic function will be expressed in terms
 of $\omega$, the logarithm of a (holomorphic) modular form that was
 identified in \cite{Gregori:1999ns}, and which depends on either one
 of the three complex moduli, $S$, $T$, or $U$, and its multiple
 holomorphic derivatives.  At lowest non-trivial order duality
 invariance is achieved by allowing for the presence of a term
 $\ln X^0$ which, strictly speaking, is not part of the Wilsonian
 effective action. It is not unique, as there are non-holomorphic
 alternatives, and it has appeared at various stages in the
 literature.  The fields $X^0$, $\mathrm{i}X^0S$, $\mathrm{i}X^0T$ and
 $\mathrm{i}X^0U$ are the complex scalar fields belonging to the four
 off-shell vector multiplets of the underlying supergravity.

 From the holomorphic function and its complex conjugate one can
 derive a version of the topological string partition function of the
 STU-model by following the procedure outlined in
 \cite{Cardoso:2014kwa}.\footnote{
   Actualy, the present study of this STU-model was partially
   motivated by the need to put the results of \cite{Cardoso:2014kwa}
   to a test in the context of a realistic model.}
 This amounts to constructing the corresponding (real) Hesse potential
 by means of a suitable Legendre transformation. The Hesse potential
 depends on duality covariant moduli and it transforms as a function
 under general duality transformations, which for the STU-model
 constitute the group $\mathrm{Sp}(8;\mathbb{R})$. Consistency
 therefore requires that it is invariant under its $[\Gamma_0(2)]^3$
 subgroup and under triality. The full Hesse potential is a real
 function, but as was demonstrated in \cite{Cardoso:2014kwa}, it
 decomposes into an infinite number of separately invariant functions
 of which precisely one exhibits the characteristic features of a
 topological string partition function. For instance, it is harmonic
 in the holomorphic modular form $\omega$, so that it will decompose
 into a sum of a function of $\omega$ and one of $\bar\omega$, but
 both these functions will still depend on the moduli and on their
 complex conjugates. The lack of homorphicity can be characterized in
 terms of of a holomorphic anomaly equation, something that is a
 well-known feature of the topological string
 \cite{Bershadsky:1993ta,Bershadsky:1993cx}. The holomorphic anomaly
 equation was generally derived in \cite{Cardoso:2014kwa} on the basis
 of the diagrammatic structure in the Hesse potential. Upon including
 certain non-holomorphic terms for the genus-1 partition function, it
 was established that this anomaly equation belongs to the same class
 as the one that is known for the topological
 string.\footnote{ 
   It has been shown meanwhile that the holomorphic anomaly equation
   has an interpretation as an integrability condition for the
   existence of a Hessian structure \cite{Cardoso:2015qhq}.} 
 By explicit calculations one can then identify the leading terms with
 the low-genus topological string partition functions of the
 STU-model.

Subsequently we shall attempt to derive an all-order result for this
topological string partition function. Here we follow an approach
inspired by the work of \cite{Alim:2015qma}, which enables us to resum
a subclass of non-holomorphic terms by making use of the holomorphic
anomaly equation. The result, which turns out to be qualitatively
different from the result obtained in \cite{Alim:2015qma}, can be
further generalized by making it consistent with duality. This leads
to a modified effective coupling constant $u$, which is duality invariant
and depends on the moduli $S$, $T$, $U$. As it turns out, this
effective coupling constant takes its values on a Riemann surface
$\mathbb{C}$ .

However, this result does not yet cover the full result for the
topological string, because certain terms that have already been
derived at low orders in perturbation theory, are not
contained in the resummation. These terms are separately duality
invariant and the obvious question is whether one can also extend them
to all orders. As we will demonstrate, the coefficient functions,
which depend on $u$, of terms that are products of three identical
structures, whose arguments are equal to $S$, $T$, and $U$,
respectively, satisfy differential equations that will lead to
integration constants. Coefficient function of terms that are not of
this type can be determined algebraically. However, the integration
constants are directly related to specific terms in the holomorphic
function that has been determined earlier. Given that the
construction of this function can in principle be continued to all
orders, it seems that the dualities of the STU-model indeed determine
the holomorphic function and the corresponding topological string
partition function.

This paper is organized as follows. In section
\ref{sec:STU-model-dualities} we introduce the main features of the
STU-model and its dualities as well as the consequences of the
STU-dualities for the holomorphic function, whose lowest-order
contribution involves the logarithm of $X^0$.  Various features of
this logarithmic term as well as alternative versions are
discussed. Section \ref{sec:higher-orders} describes the results for
the holomorphic function up to fifth order in perturbation theory,
which are obtained by imposing invariance under the dualities. We are
not aware of any possible impediment for continuing this strategy to
arbitrary orders of perturbation theory, and we will assume that this
approach can in principle be continued to any given order.

Section \ref{sec:topol-string-part} describes how to obtain the
lowest-order results for a corresponding version of the topological string
partition function, as well as its holomorphic anomaly equation, by
following the method of \cite{Cardoso:2014kwa}. This requires a
Legendre transform, that was performed iteratively, leading to the
so-called Hesse potential that is a function of duality covariant
variables. One particular subsector takes the form of a topological
string partition function and transforms as a function under duality
transformations. It depends holomorphically on the topological string
coupling, with multiple covariant derivatives of the holomorphic
modular forms $\omega(S)$, $\omega(T)$, and $\omega(U)$. However,
these covariant derivatives contain non-holomorphic connections, so
that the result is not holomorphic in the moduli $S$, $T$, and $U$.

In section \ref{sec:partial-det-h}, we start the derivation of an
all-order result for the topological string partition function. We
first present a derivation of a function that captures all the
non-holomorphic features of the topological string partition function,
following an approach inspired by the work of
\cite{Alim:2015qma}. After covariantizing this function one may
collect the remaining terms into a second invariant function, whose
leading terms in perturbation theory are now known and have a
systematic structure in terms of a set of duality invariants.

Section \ref{sec:addit-contr-h1} is then devoted to the determination
of this last function, by imposing the holomorphic anomaly
equation. Here we note that there are infinitely many different
structures, as we are dealing with an infinite variety of
invariants. Exploring the various terms we find that most of them can
be determined algebraically in this way, while the remaining ones are
subject to differential equations, which can be solved up to
integration constants. Nevertheless it turns out that these
integration constants can still be fixed by a careful comparison
between the results for the topological string and those for the
holomorphic function that encodes the Wilsonian effective action. 
This is one more indication that both the Wilsonian action and the
topological string can be uniquely determined to all orders by
requiring that the dualities act consistently.

The final section \ref{sec:conclusions} presents our conclusions. We
enclose a brief appendix \ref{sec:formulae} that contains a few useful
formulae and a second appendix \ref{sec:Large-moduli-limits} in which
we analyze the results in the limit where the real parts of two of the
moduli are taken to infinity. In this case there are substantial
simplifications.

\section{The STU-model and its dualities}
\label{sec:STU-model-dualities}
\setcounter{equation}{0}

The effective action for the massless modes of the STU-model of Sen
and Vafa can be described in terms of $N=2$ supergravity coupled to
$n_\mathrm{v}=3$ vector multiplets and $n_\mathrm{h}=4$
hypermultiplets, so that the Euler characteristic
$\chi\equiv 2( n_\mathrm{v}-n_\mathrm{h}+1)$ will vanish, as is
required by the fact that the type-II description should be based on a
self-mirror Calabi-Yau manifold.  At the classical level its moduli
space can be written as a product of the following special-K\"ahler
and quaternion-K\"ahler spaces,
\begin{equation}
  \label{eq:SK+QK-spaces}
  \mathcal{M}_\mathrm{vector} =
  \frac{\mathrm{SL}(2)}{\mathrm{SO}(2)} \times
  \frac{\mathrm{SL}(2)}{\mathrm{SO}(2)}\times
  \frac{\mathrm{SL}(2)}{\mathrm{SO}(2)}  \,,\qquad 
    \mathcal{M}_\mathrm{hyper} = 
    \frac{\mathrm{SO}(4,4)}{\mathrm{SO}(4)\times \mathrm{SO}(4) }\,. 
\end{equation}
In what follows we will focus on the vector multiplet sector.

The off-shell effective action for the massless fields is described in
terms of four vector multiplets which contain four vector gauge
fields, $W_\mu{\!}^0$, $W_\mu{\!}^1$, $W_\mu{\!}^2$ and $W_\mu{\!}^3$,
as well as four complex scalars $X^0$, $X^1$, $X^2$, $X^3$. This
description is locally superconformally invariant and therefore these
fields are subject to local dilatations and phase transformations. As
is well known, the Wilsonian effective action is encoded in a
holomorphic function of these scalar fields that is homogeneous of
second degree under complex scale transformations
\cite{deWit:1984pk}. At the classical level this function takes the form
\begin{equation}
  \label{eq:class-F}
  F(X) = -\, \frac{X^1\, X^2\,X^3}{X^0} \,,
\end{equation} 
whose corresponding supergravity action leads precisely to the
special-K\"ahler moduli space specified in \eqref{eq:SK+QK-spaces}.
The isometry group of this moduli space is the direct product of three
independent $\mathrm{SL}(2)$ groups, and not the product of their
respective $\Gamma_0(2)$ subgroups. However, the function $F(X)$ will
also contain terms that describe gravitational couplings of the
special-K\"ahler moduli, which are only invariant under the direct
product of the independent $\Gamma_0(2)$ groups \cite{Gregori:1999ns}.
Upon modding out the special-K\"ahler space in \eqref{eq:SK+QK-spaces}
by the action of the integer-valued group $[\Gamma_0(2)]^3$, the
combined action will then be invariant under the latter
group. Moreover, the invariance under permutations of the fields
$X^1$, $X^2$ and $X^3$, known as {\it triality}, is respected by the
function \eqref{eq:class-F} as well as by its modification that we
will introduce shortly.

The relevant special-K\"ahler moduli are conventionally denoted by
$S$, $T$ and $U$, and defined by
\begin{equation}
  \label{eq:def-STU}
  S=-\mathrm{i} \,\frac{X^1}{X^0} \,,\qquad T=-\mathrm{i}
  \,\frac{X^2}{X^0} \,,\qquad U=-\mathrm{i} \,\frac{X^3}{X^0} \,.  
\end{equation}
These fields parametrize the special-K\"ahler target space and they
are invariant under local dilatations and phase transformations. Since we
intend to remain in the off-shell formulation we will retain the field
$X^0$.

Since the supergravity description contains four vector gauge fields,
one belonging to the Poincar\'e supergravity multiplet and one for
each of the three matter multiplets, one will also be dealing with four
magnetic charges denoted by $p^0$, $p^1$, $p^2$, $p^3$, and four
electric charges denoted by $q_0$, $q_1$, $q_2$, $q_3$. These charges
are carried by the underlying microscopic degrees of freedom of the
STU-model. Under S-duality they
transform as follows,
\begin{equation}
    \label{eq:S-charge-duality}
    \begin{array}{rcl}
      p^0 &\!\to\!& d \, p^0 + c\, p^1 \;,\\
      p^1 &\!\to\!& a \, p^1 + b \, p^0 \;,\\
      p^2 &\!\to\!& d\, p^2 - c \,q_3  \;,\\
      p^3 &\!\to\!& d\, p^3 - c \,q_2  \;,
    \end{array}
    \qquad
    \begin{array}{rcl}
      q_0 &\!\to\!& a\,  q_0 -b\,q_1 \;, \\
      q_1 &\!\to\!& d \,q_1 -c\, q_0 \;,\\
      q_2 &\!\to\!& a \, q_2 -b\, p^3 \;,\\
      q_3 &\!\to\!& a\, q_3 - b\, p^2  \;. 
    \end{array}
\end{equation}
In the  STU-model the charges should take their values in an
eight-dimensional discrete lattice that will only be invariant under
the action of the
$\Gamma_0(2)_\mathrm{S}\times\Gamma_0(2)_\mathrm{T}
\times\Gamma_0(2)_\mathrm{U}$ duality group, so that the parameters
$a$, $b$, $c$, and $d$ must be restricted accordingly.  Based on the
function \eqref{eq:class-F}, the moduli fields transform under 
$\Gamma_0(2)_\mathrm{S}$ as 
\begin{equation}
  \label{eq:STU-S-dual}
  S\to \frac{a\,S-\mathrm{i} b}{d+ \mathrm{i}c\,S} \,,\qquad T\to
  T\,,\qquad U\to U\,, \qquad X^0\to  (d+ \mathrm{i}c\,S) \,X^0 \,. 
\end{equation}
Similar results apply to T- and U-duality transformations, which are
directly obtained upon interchanging the labels $1\leftrightarrow2$
(or $1\leftrightarrow3$) and correspondingly $S\leftrightarrow T$ (or
$S\leftrightarrow U$).  From these transformation rules it follows that the
eight charges will transform according to the
$(\mathbf{2},\mathbf{2},\mathbf{2})$ representation of
$\Gamma_0(2)_\mathrm{S}\times\Gamma_0(2)_\mathrm{T}
\times\Gamma_0(2)_\mathrm{U}$.

As mentioned above, the function \eqref{eq:class-F} will contain
additional terms that break the original $[\mathrm{SL}(2)]^3$
dualities to the subgroup
$\Gamma_0(2)_\mathrm{S}\times\Gamma_0(2)_\mathrm{T}
\times\Gamma_0(2)_\mathrm{U}$. This must be done is such a way that
the action of the duality subgroup on the field strengths and their
duals will be preserved. Furthermore, as was already mentioned, the
Wilsonian supersymmetric effective action for the STU-model must be
encoded in a holomorphic function that is homogeneous of degree two
\cite{deWit:1984pk}, and should be consistent with the dualities as
well as with triality.  As it turns out, this puts stringent
constraints on the way in which we can modify the function
\eqref{eq:class-F}.  To see this we start from the following
holomorphic function,
\begin{equation}
  \label{eq:holo-function}
  F(X,A) = -\,\frac{X^1X^2X^3}{X^0} +2\mathrm{i}\,\Omega(X,A)\,,  
\end{equation}
where the first term describes the Lagrangian that is at most
quadratic in space-time derivatives of the vector-multiplet
fields. The extra term $\Omega$ contains a holomorphic parameter $A$,
which actually corresponds to a field. This field $A$ is the lowest
component of the square of a tensor chiral supermultiplet, known as
the Weyl multiplet. Its presence in \eqref{eq:holo-function} will lead
to higher-derivative interactions that involve among others the square
of the Weyl tensor. Supersymmetry requires the function $F$ to be
homogenous of degree two, i.e.
$F(\lambda X, \lambda^2 A) = \lambda^2 F(X,A)$. The Wilsonian action
will therefore be based on such a homogeneous holomorphic
function.\footnote{
  We should point out that alternative higher-derivative couplings
  exist for these models, but they are not holomorphic
  \cite{deWit:2010za, Butter:2013lta}. Their behaviour under
  electric-magnetic duality has so far not been investigated in much
  detail and they do not contribute to the Wilsonian effective action
  nor to the topological string partition function.}

The duality transformations such as \eqref{eq:S-charge-duality} are
generated on the fields $X^I$ by electric-magnetic duality and this
ensures that they will also act accordingly on the electric and
magnetic charges. The Lagrangian and its underlying function $F(X,A)$
are therefore not invariant under the duality transformations. Rather
the requirement is that the `periods' $(X^I, F_J(X))$ must transform
covariantly under the dualities \cite{deWit:1984pk,Cecotti:1988qn},
precisely as the charges $(p^I,q_J)$ (see
\eqref{eq:S-charge-duality}). Hence the duality transformations
involve the holomorphic derivatives of $F(X,A)$ which we list here for
convenience (we refrain from indicating the dependence on $A$ for
notational simplicity),
\begin{align}  
  \label{eq:F_I-STU}
  F_0(X)=&\; \frac{X^1X^2X^3}{(X^0)^2} -\frac{2\mathrm{i}}{X^0}
  \left[-X^0 \frac{\partial}{\partial X^0}+
  S\frac{\partial}{\partial S}+ T\frac{\partial}{\partial T} +
  U\frac{\partial}{\partial U} 
  \right]\Omega \;,  \nonumber\\ 
  F_1(X)=&\; - \frac{X^2X^3}{X^0} +\frac{2}{X^0} \,
  \frac{\partial\Omega}{\partial S} \;,\nonumber\\
  F_2(X) =&\; - \frac{X^1X^3}{X^0} +\frac{2}{X^0} \,
  \frac{\partial\Omega}{\partial T} \;, \nonumber\\
  F_3(X) =&\; - \frac{X^1X^2}{X^0} +\frac{2}{X^0} \,
  \frac{\partial\Omega}{\partial U} \;.
\end{align}
The above formulae clearly exhibit the triality symmetry, provided
that $\Omega$ is triality invariant. The field $A$ is not subject to
the duality transformations. 

To construct the duality transformations on the fields, one considers
the S-dualities acting on the charges $p^I$ and $q_I$ given in
\eqref{eq:S-charge-duality} and apply the same transformations on the
periods $X^I$ and $F_I$. The fields will thus transform as follows,
\begin{equation}
  \label{eq:full-S-STU}
  \begin{array}{rcl}
  X^0&\!\!\to\!\!& X^0{}'= \Delta_{\mathrm{S}}\, X^0\;,\\[.4ex]
  X^2&\!\!\to\!\!& X^2{}'= \Delta_{\mathrm{S}}\, X^2 - 
                   \displaystyle{\frac{2\, c}{X^0} \,
  \frac{\partial\Omega}{\partial U}}  \;, 
  \end{array}
  \qquad 
  \begin{array}{rcl}
  X^1&\!\!\to\!\!& X^1{}' =  a\,X^1+ b\,X^0\;, \\[.4ex]
  X^3&\!\!\to\!\!& X^3{}'= \Delta_{\mathrm{S}}\, X^3 -
                   \displaystyle{{\frac{2\,c}{X^0} \, 
  \frac{\partial\Omega}{\partial T}}}  \;,  
  \end{array}
\end{equation}
where $a,b,c,d$ refer to the parameters of the S-duality
transformation and $\Delta_\mathrm{S}(S)$ is defined by 
\begin{equation}
  \label{eq:def-Delta}
  \Delta_\mathrm{S}(S)= d+ \mathrm{i}c\,S\,. 
\end{equation}
Observe that there exist similar quantities $\Delta_\mathrm{T}(T)$ and
$\Delta_\mathrm{U}(U)$ with parameters $d$ and $c$ that belong to the
T- and U-duality transformations, respectively.  Furthermore we note
the convenient relations
\begin{equation}
  \label{eq:extra-dual-rel}
  \frac{\partial S'}{\partial S} = \frac{1}{\Delta_\mathrm{S}{}^{2}}\;,\qquad
  \frac1{S+\bar S} \to \frac{\vert \Delta_\mathrm{S}\vert^2} {S+\bar S}
  =  \frac{\Delta_\mathrm{S}{}^2} {S+\bar S} -
  \Delta_\mathrm{S}\,\frac{\partial\Delta_\mathrm{S}}{\partial S} \,. 
\end{equation}
The results lead to the following transformations of $X^0$, $S$,
$T$ and $U$,
\begin{align}
  \label{eq:STUY0-full}
  X^0\rightarrow&\; X^{0\,\prime} = \Delta_\mathrm{S} \,X^0\;,
                  \nonumber\\[1mm] 
  S \rightarrow&\; S' = 
  \frac{a\,S-\mathrm{i}b}{\Delta_\mathrm{S}} \;, \nonumber\\[1mm]
  T\rightarrow&\; T' = T +\frac{2}{\Delta_{\mathrm{S}}\,(X^0)^2}
  \,\frac{\partial\Delta_\mathrm{S}}{\partial S}
  \,\frac{\partial\Omega}{\partial U}  \;, \nonumber\\[1mm] 
  U\rightarrow&\;  U' =U
  +\frac{2}{\Delta_{\mathrm{S}}\,(X^0)^2}\,
   \frac{\partial\Delta_\mathrm{S}}{\partial S} 
  \,\frac{\partial\Omega}{\partial T}  \;,
\end{align}
so that the S-duality transformations on $X^0$ and $S$ remain
unchanged, while the fields $T$ and $U$ will now transform
non-trivially. Obviously the T- and U-duality transformations follow
from triality.

Because the STU-dualities are assumed to define symmetries of the
model to all orders in perturbation theory, the transformation of the
derivatives \eqref{eq:F_I-STU} must coincide with the results obtained
by explicitly substituting the transformed values of the moduli given
by \eqref{eq:STUY0-full} into the expressions for $F_I$,
\begin{align}
  \label{eq:F-I-S-dual}
  F_0(X') =&\; a\, F_0(X) - b \, F_1(X)\,,\nonumber\\
  F_1(X') =&\; d\, F_1(X) - c \, F_0(X)\,,\nonumber\\
  F_2(X') =&\; a\, F_2(X) - b \, X^3 \,,\nonumber\\
  F_3(X') =&\; a\, F_3(X) - b \, X^2  \,.
\end{align}
In that case the periods $(X^I, F_J(X))$ transform covariantly under
the dualities precisely as the charges $(p^I,q_J)$, as
shown in \eqref{eq:S-charge-duality}. Note again that the function $F(X)$
itself does not transform as a function under electric-magnetic
duality, which explains the precise form of the left-hand side of
\eqref{eq:F-I-S-dual}. The above equations \eqref{eq:F-I-S-dual} lead
to conditions on the derivatives of $\Omega$ that take the following
form,
\begin{align}
  \label{eq:S-invariance-STU}
  \bigg(\frac{\partial\Omega}{\partial T}\bigg)^\prime_\mathrm{S} =&\;
  \frac{\partial\Omega}{\partial T} \;,\qquad\quad
  \left(\frac{\partial\Omega}{\partial U}\right)^\prime_\mathrm{S} =~
  \frac{\partial\Omega}{\partial U} \;,
 \nonumber\\[1mm]
  \bigg(\frac{\partial\Omega}{\partial S}\bigg)^\prime_\mathrm{S} -
  \Delta_{\mathrm{S}}{\!}^2\,\frac{\partial\Omega}{\partial S} =&\;
  \frac{\partial\Delta_{\mathrm{S}}}{\partial{S}} 
  \bigg[- \Delta_{\mathrm{S}} \, X^0\, \frac{\partial\Omega}{\partial X^0}
  -\frac{2}{(X^0)^2}\,
  \frac{\partial\Delta_{\mathrm{S}}}{\partial{S}} 
  \,\frac{\partial\Omega}{\partial T} 
  \frac{\partial\Omega}{\partial U} \bigg] \;,
  \nonumber\\[1mm] 
  \bigg(X^0\frac{\partial\Omega}{\partial X^0}\bigg)^\prime_\mathrm{S} =&\;
  X^0 \frac{\partial\Omega}{\partial X^0} 
  +\frac{4}{\Delta_{\mathrm{S}}\,(X^0)^2}
  \,\frac{\partial\Delta_{\mathrm{S}}}{\partial{S}}\, 
  \,\frac{\partial\Omega}{\partial T}
  \frac{\partial\Omega}{\partial U}\;.
\end{align}
These equations, which are non-linear in $\Omega$, were first derived
in \cite{Cardoso:2008fr} for the more general case where $\Omega$ is
not necessarily holomorphic. Note that the prime on the quantities on
the left-hand side indicates that we have replaced all the fields by
their transformed ones specified in
\eqref{eq:STUY0-full}. Corresponding results for T- and U-duality
follow directly upon applying triality. As it turns out, the above
equations are very restrictive, especially when insisting on
triality. The function $\Omega$ can then be solved by iteration in powers
of $A$, depending on some initial conditions. A noteworthy feature of
the equations \eqref{eq:S-invariance-STU} is that they depend
non-linearly on $\Omega$. The solution of these equations based on a
power series in $A$ seems to be unique.

The iteration is based on the fact that $\Omega(X,A)$ must be a
homogeneous function of second degree, which can be expanded in terms
of a auxiliary complex field $A$ which scales with weight
two. Because the fields $S,T,U$ do not scale,  $X^0$ and $A$ are the
only fields that are subject to scale transformations. Therefore
$\Omega$ can be written in a series expansion in powers of $A\,(X^0)^{-2}$
with coefficient functions that depend on $S,T,U$ and an overall
factor $A$,
\begin{equation}
  \label{eq:Ups-holo-expansion}
  \Omega(X,A) = A\, \bigg[
  \gamma\,\ln\frac{(X^0)^2}{A} + \omega^{(1)}(S,T,U) +
  \sum_{n=1}^{\infty} \, 
  \Big(\frac{A}{(X^0)^2} \Big)^{n} \, \omega^{(n+1)}(S,T,U)\bigg]
  \,.
\end{equation}
Note that we allowed for the presence of a logarithmic term, which
under S-duality transforms with a shift proportional to $\ln
\Delta_\mathrm{S}(S)$, with corresponding variations under T- and
U-duality. 

Let us now first concentrate on the lowest-order terms which should
reproduce the result of \cite{Gregori:1999ns}.  Under S-duality one
can directly analyze the equations~\eqref{eq:S-invariance-STU} in
lowest order in $A$. The last equation is trivially satisfied, whereas
the other three equations give rise to the following restrictions on
$\omega^{(1)}(S,T,U)$,
\begin{align}
  \label{eq:omega-1}
   \bigg(\frac{\partial\omega^{(1)}}{\partial T}\bigg)^\prime_\mathrm{S} =&\;
  \frac{\partial\omega^{(1)}}{\partial T} \;,\qquad\quad
  \bigg(\frac{\partial\omega^{(1)}}{\partial U}\bigg)^\prime_\mathrm{S} =~
  \frac{\partial\omega^{(1)}}{\partial U} \;,
 \nonumber\\[2mm]
  \bigg(\frac{\partial\omega^{(1)}}{\partial S}\bigg)^\prime_\mathrm{S} =&\;
  \Delta_{\mathrm{S}}(S)^2\,\frac{\partial\omega^{(1)}}{\partial S} - 2\gamma\,
  \Delta_{\mathrm{S}} \,
  \frac{\partial\Delta_{\mathrm{S}}}{\partial{S}}   \;. 
\end{align}
Upon imposing triality the combined equations show that 
\begin{equation}
  \label{eq:sol-omega-1}
  \omega^{(1)}(S,T,U)=
\omega(S)+\omega(T)+\omega(U) \,,
\end{equation}
where $\omega(S)$ must be the logarithm of a modular form which must
transform as 
\begin{equation}
  \label{eq:S-var-omega}
  \omega(S') = \omega(S)- 2\gamma \,\ln \Delta_\mathrm{S}(S)\,. 
\end{equation}
Here we note that the transformation of the function $\omega(S)$ is in
principle not fully captured by \eqref{eq:S-var-omega}, because the
right-hand side may also include a constant imaginary shift due to the
multiplyer system of the modular form. Such a shift, if present, is
harmless.  The reason is that an imaginary constant shift leads to a
real constant times the imaginary part of $A$ in the effective
action. However, the imaginary part of $A$ encodes a Lagrangian that
equals a total derivative, which can be ignored. Beyond the lowest
order we will only be dealing with derivatives of $\omega(S)$, so that
this imaginary shift is no longer relevant.

Obviously similar results hold for $\omega(T)$ and $\omega(U)$
by triality. Upon comparing \eqref{eq:S-var-omega} to the result found
in \cite{Gregori:1999ns} for the one-loop gravitational coupling in the
STU-model, we must choose $\omega(S)$ equal to 
\begin{equation}
  \label{eq:explict-omega}
  \omega(S) = \frac1{64 \pi}\,  \ln \vartheta_2(S) \,, \quad
  \mathrm{with} \quad
  \gamma = -\frac1{256\pi}\,,
\end{equation}
whose corresponding duality group is precisely
$\Gamma_0(2)$.\footnote{ 
  We note that while the expression for $ \omega^{(1)}(S,T,U)$ given
  in \eqref{eq:sol-omega-1} has manifest triality symmetry, the
  corresponding expression given in equation (2.6) of
  \cite{Gregori:1999ns} involves both $ \ln \vartheta_2$ and
  $\ln \vartheta_4$. However, by applying the modular transformation
  $\tau \rightarrow - 1/\tau$ to $\ln \vartheta_4$, the expression
  given in \cite{Gregori:1999ns} becomes manifestly triality
  symmetric, and it agrees with the expression for
  $ \omega^{(1)}(S,T,U)$ given in \eqref{eq:sol-omega-1}.}
We note that $\vartheta_2(S)$ can be expressed in terms of the
Dedekind function as $\vartheta_2(S)= 2\, \eta^2(2S)/\eta(S)$.  Hence
the choice \eqref{eq:explict-omega} restricts the STU-dualities to the
$\Gamma_0(2)$ subgroups of the generic $\mathrm{SL}(2)$ duality
groups, so that the matrix elements $a$, $b$, $c$, $d$ should satisfy
the restrictions pointed out in section \ref{sec:introduction}.

The fact that $\gamma$ is different from zero is important for the
iteration that will be performed in the next section. In this
iteration the quantity $A/(X^0)^2$ will play the role of a coupling
`constant', which is invariant under local dilatations and phase
transformations, but not under the dualities. In
the result of \cite{Gregori:1999ns}, the field $X^0$ is not present,
so the shift in the variation of $\omega(S)$ has to be cancelled by
some other term, such as
\begin{align}
  \label{eq:non-holo-threshold}
  -2 \gamma \ln [(S+\bar S)(T+\bar T)(U+\bar U)]
  \stackrel{\mathrm{S}} {\longrightarrow} &\, -2\gamma \ln
          [(S+\bar S)(T+\bar T)(U+\bar U)] \nonumber\\
        &\, + 2\gamma \ln \Delta_\mathrm{S}(S)
          + 2\gamma \ln {\bar\Delta}_\mathrm{S}(\bar{S}) \,, 
\end{align}
and likewise for T- and U-duality. Obviously these variations are
identical to those of $\gamma [ \ln [(X^0)^2/A] + \text{h.c.}]$.
However, within the context of the effective action the variation
\eqref{eq:non-holo-threshold} is only an approximation, because the
transformations of the moduli are corrected in view of
\eqref{eq:STUY0-full}. Furthermore it does not make sense to add a
term to the effective action that explicitly involves the moduli,
whose duality transformations are actually governed by the period
vector $(X^I,F_J)$, where $F_J$ is defined as the derivative of the
full function $F(X)$ with respect to $X^J$.

The above situation is, however, not problematic, because the big
moduli space that involves the field $X^0$ is still subject to local
dilatations as well as $\mathrm{U}(1)$ gauge transformations and there
exists a gauge condition that replaces $\ln \vert X^0\vert$ in terms
of a non-holomorphic term whose duality transformation agrees in
leading order with \eqref{eq:non-holo-threshold}. To see this we
introduce a generalized K\"ahler potential $\mathcal{K}$, defined by
\begin{equation}
  \label{eq:rel-k-ln-sympl}
  \mathcal{K} - \ln \vert X^0\vert^2  = -\ln \big[\mathrm{i}
  \bar X^I \, F_I(X)  - \mathrm{i} \bar F_I(\bar  X)\,  X^I \big] \,, 
\end{equation}
where the right-hand side is manifestly duality invariant and
transforms under local dilatations.  If we now impose a gauge
condition for local dilatations by constraining
$\big[\mathrm{i} \bar X^I \, F_I(X) - \mathrm{i} \bar F_I(\bar X)\,
X^I \big]$ to a real constant, then $\ln\vert X^0\vert$ is equal to a
non-holomorphic expression whose leading contribution coincides with
\eqref{eq:non-holo-threshold},
\begin{align}
  \label{eq:K-classical}
  \mathcal{K} =&\,  - \ln [(S+\bar
  S) (T+\bar T) (U+\bar U)]+ \mathcal{O} (\Omega) \,. 
\end{align}
Obviously it is convenient to work with the field $X^0$ throughout the
calculations and to postpone imposing this gauge choice until the
end.\footnote{
  It is worth pointing out that there is another non-holomorphic
  quantity that depends only on the periods that is invariant under
  dilatations and $\mathrm{U}(1)$ transformations, namely the metric
  \begin{equation}
  \label{eq:def-N}
  N_{IJ}(X,\bar X)\equiv 2\,\mathrm{Im} \,[F_{IJ}(X)] \,. 
\end{equation}
A special feature of the STU-model is that
$\ln \vert \det N_{IJ}\vert$ satisfies a similar limit as
$\mathcal{K}$, namely 
\begin{equation}
  \label{eq:ln-N}
  \ln \vert \det N_{IJ}\vert=2 \ln [(S+\bar S) 
  (T+\bar T) (U+\bar U)]+ \mathcal{O} (\Omega) \,. 
\end{equation}
 }  

We should add that the logarithmic term $\ln X^0$ has appeared at
various stages in the literature in the study of BPS black hole
entropy. It was first discussed in \cite{Dabholkar:2005dt} using the
Gopakumar-Vafa term \cite{Gopakumar:1998ii,Gopakumar:1998jq} when
calculating black hole entropy corrections. At the black hole horizon
$A/(X^0)^2$ is inversely proportional to the square of the charges as
a result of the BPS attractor equations
\cite{LopesCardoso:2000qm}. The factor appears in the measure of an
OSV-type integral derived in \cite{Denef:2007vg} for BPS black
holes. The presence of this term is moreover consistent with the
result of the logarithmic corrections to black hole entropy
\cite{Sen:2011ba}.  We refer to section~4.2.4 in
\cite{Dedushenko:2014nya} for further comments regarding this
logarithmic term and its possible origin.

Before moving to the next section we discuss some technical aspects
regarding the duality transformations of derivatives of $\omega(S)$
under S-duality. We list the first few multiple derivatives,
\begin{align} 
  \label{eq:transf-higher-der-omega}
  \frac{\partial\omega}{\partial S} \to&\; {\Delta_\mathrm{S}}^2 \,
  \frac{\partial\omega}{\partial S} - 2\gamma\, \Delta_\mathrm{S} \,
  \frac{\partial\Delta_\mathrm{S}}{\partial S} \,, \nonumber\\[1mm]
  \frac{\partial^2\omega}{\partial S^2} \to&\; {\Delta_\mathrm{S}}^4 \,
  \frac{\partial^2\omega}{\partial S^2} + 2\, {\Delta_\mathrm{S}}^3 \,
  \frac{\partial\Delta_\mathrm{S}}{\partial S} \,\frac{\partial\omega}{\partial
    S} - 
  2\gamma\, {\Delta_\mathrm{S}}^2 \,
  \Big(\frac{\partial\Delta_\mathrm{S}}{\partial S}\Big)^2 \,, \nonumber\\[1mm]
 \frac{\partial^3\omega}{\partial S^3} \to&\; {\Delta_\mathrm{S}}^6 \,
  \frac{\partial^3\omega}{\partial S^3} + 6\, {\Delta_\mathrm{S}}^5 \,
  \frac{\partial\Delta_\mathrm{S}}{\partial S} \,\frac{\partial^2\omega}{\partial
    S^2}  + 6\, {\Delta_\mathrm{S}}^4 \,
  \Big(\frac{\partial\Delta_\mathrm{S}}{\partial S} \Big)^2\,\frac{\partial\omega}{\partial
    S}  - 
  4\gamma\, {\Delta_\mathrm{S}}^3 \,
  \Big(\frac{\partial\Delta_\mathrm{S}}{\partial S}\Big)^3 \,, \nonumber\\[1mm]
 \frac{\partial^4\omega}{\partial S^4} \to&\; {\Delta_\mathrm{S}}^8 \,
  \frac{\partial^4\omega}{\partial S^4} + 12\, {\Delta_\mathrm{S}}^7 \,
  \frac{\partial\Delta_\mathrm{S}}{\partial S} \,\frac{\partial^3\omega}{\partial
    S^3} + 36\, {\Delta_\mathrm{S}}^6 \,
  \Big(\frac{\partial\Delta_\mathrm{S}}{\partial S} \Big)^2\,\frac{\partial^2\omega}{\partial
    S^2}  \nonumber \\
  &\: + 24\, {\Delta_\mathrm{S}}^5 \,
  \Big(\frac{\partial\Delta_\mathrm{S}}{\partial S}
  \Big)^3\,\frac{\partial\omega}{\partial  S}  - 
  12\gamma\, {\Delta_\mathrm{S}}^4 \,
  \Big(\frac{\partial\Delta_\mathrm{S}}{\partial S}\Big)^4 \,.
\end{align} 
The presence of the derivatives on $\Delta(S)$ on the right-hand side
indicates that we are not dealing with covariant quantities. Therefore
we explicitly modify the derivatives on the left-hand side to
eliminate these non-covariant variations, and denote the resulting
covariant expressions by $I^{(2)}(S)$, $I^{(3)}(S)$ and $I^{(4)}(S)$,
which will transform as
\begin{equation}
  \label{eq:higher-order procucts} 
  I^{(n)}(S)  \to  I^{(n)}(S') =   \Delta_\mathrm{S}(S)^{2n} \,  I^{(n)}(S)\,. 
\end{equation}
The explicit expressions are
\begin{align}
   \label{eq:def-I}
  I^{(2)}(S) =&\,  \frac{\partial^2\omega}{\partial S^2} +\frac1{2\gamma} \,
  \Big(\frac{\partial\omega}{\partial S}\Big)^2\,, \nonumber\\[2mm]
  I^{(3)} (S)=&\, \frac{\partial^3\omega}{\partial S^3} +\frac3{\gamma}\,
    \frac{\partial^2\omega}{\partial S^2}
    \,\frac{\partial\omega}{\partial S} + \frac1{\gamma^2}\,
  \Big(\frac{\partial\omega}{\partial S}\Big)^3\,,\nonumber\\[2mm]
  I^{(4)}(S) =&\, \frac{\partial^4\omega}{\partial S^4} +\frac6{\gamma}\,
    \frac{\partial^3\omega}{\partial S^3} 
    \,\frac{\partial\omega}{\partial S} +\frac{3}{\gamma} \,
    \Big(\frac{\partial^2\omega}{\partial S^2}\Big)^2  +
    \frac{12}{\gamma^2}\,  \frac{\partial^2\omega}{\partial S^2} 
    \,\Big(\frac{\partial\omega}{\partial S}\Big)^2 
    + \frac{3}{\gamma^3}\,\Big(\frac{\partial\omega}{\partial
      S}\Big)^4\,. 
 \end{align}

 Because products of the $I^{(n)}(S)$ will also transform covariantly,
 the explicit expressions $I^{(n)}$ with $n>3$ are in principle
 ambiguous. For the expressions above we made sure that the relation
\begin{equation}
  \label{eq:DI-I}
  I^{(n+1)}(S)  = \mathcal{D}_S  I^{(n)}(S)\,, 
\end{equation}
holds, by making a specific choice for $I^{(4)}$. Here $\mathcal{D}_S$
denotes a holomorphic covariant derivative, known as the Serre
derivative (for more details, see \cite{Hahn:2015}),
which acts on $I^{(n)}$ as
\begin{equation}
  \label{eq:serre-derivative}
  \mathcal{D}_S\, I^{(n)}(S)  \equiv \Big(\frac{\partial}{\partial S} +
  \frac{n}{\gamma}\, \frac{\partial \omega(S)}{\partial S}\Big)
  \,  I^{(n)}(S) \,. 
\end{equation}
Henceforth we will assume that \eqref{eq:DI-I} will extend to all
integer values of $n\geq2$, so as to provide a unique basis for all
the covariant expressions as polynomials in terms of the $I^{(n)}(S)$.
We note that the $I^{(n)}$ can be expressed in terms of linear
combinations of products of Eisenstein series of $\Gamma_0(2)$
\cite{Cardoso:2019avb},
\begin{equation}
  \label{eisenstein}
  I^{(n)} = \sum_{k + 2l = n\,, k \geq  0, l \geq 1} \!\!  a_{k,l}
  \;  (\tilde{\cal E}_2)^k \,  ({\cal E}_4)^l  \;,
\end{equation}
with real positive constants $a_{k,l} $.  Here, ${\cal E}_4$ is a
normalized Eisenstein series of weight $4$ of $\Gamma_0(2)$, while
$\tilde{{\cal E}}_2$ is the modular form of weight $2$ of
$\Gamma_0(2)$ given by \cite{Hahn:2015}
\begin{equation}
\tilde{{\cal E}}_2  (\tau) =  \tfrac12 \left( 3\,  {\cal E}_2 (\tau) -  E_2 (\tau) \right) = 
2 E_2(2 \tau) - E_2 (\tau) \;,
\end{equation}
where $ {\cal E}_2 (\tau), E_2(\tau)$ denote the normalized Eisenstein series of weight
$2$ of $\Gamma_0(2)$  and ${\rm SL}(2; \mathbb{Z})$, respectively.

\section{Higher-order contributions to the Wilsonian action}
\label{sec:higher-orders}
\setcounter{equation}{0}
Having determined the lowest-order result we can now proceed and
determine some higher-order contributions in the solution of
\eqref{eq:S-invariance-STU}, making use of \eqref{eq:STUY0-full}.
More precisely we will present the solutions for $\omega^{(n)}(S,T,U)$
for $n=2,3, 4,5$. Before doing so we first present the relevant
expansions in terms of a parameter $\lambda$, defined as
\begin{equation}
  \label{eq:defe-lambda}
  \lambda=\frac{A} {(X^0)^2}\,.
\end{equation}
The following results then follow straightforwardly from
\eqref{eq:Ups-holo-expansion} and \eqref{eq:STUY0-full},
\begin{align}
  \label{eq:expansions} 
  X^0\frac{\partial\Omega}{\partial X^0} =&\; A \bigg[2\,\gamma
  -2 \sum_{n=1}^\infty \lambda^{n} \,
  n\,\omega^{(n+1)}(S,T,U)\bigg] \,,\nonumber\\[1mm]
  \frac{\partial \Omega}{\partial S} =&\; A \bigg[ \frac{\partial
    \omega(S)}{\partial S} + \sum_{n=1}^\infty
     \lambda^{n} \,
  \frac{\partial\omega^{(n+1)}(S,T,U)}{\partial S} \bigg] \,,\nonumber\\[1mm]
  T \stackrel{\mathrm{S}} {\rightarrow}
   &\;T'= T + \frac{2}{\Delta_\mathrm{S}}
  \,\frac{\partial\Delta_\mathrm{S}}{\partial S} \,\bigg[
  \lambda \, \frac{\partial\omega(U)}{\partial U} +
  \sum_{n=2}^\infty \lambda^{n}
  \,\frac{\partial\omega^{(n)}(S,T,U)}{\partial U}  \bigg] \,,
\end{align}
where the last equation specifies the variation of $T$ under
S-duality. The S-duality transformation of $U$ follows from this
equation upon interchanging $T\leftrightarrow U$, whereas the
S-duality transformation of the fields $S$ and $X^0$ do not take the
form of power series, as is shown in \eqref{eq:STUY0-full}. The
transformations under T- and U-duality follow from triality.

The above expansions can now be substituted into the four equations
\eqref{eq:S-invariance-STU}. First we consider the first three
equations, where the third equation has been simplified by making use
of the fourth equation to remove the term proportional to
$(\partial\Omega/\partial T)\, (\partial\Omega/\partial U)$,
\begin{align}
  \label{eq:expansions-1}
  &\Big(\frac{\partial \omega(T)}{\partial T}\Big)^\prime_\mathrm{S}  -
  \frac{\partial \omega(T)}{\partial T}  
   +\sum_{n=1}^\infty \lambda^n \bigg[ \frac1{\Delta_\mathrm{S}{}^{2n}}
    \,\Big(\frac{\partial \omega^{(n+1)}}{\partial
      T}\Big)^\prime_\mathrm{S} - \frac{\partial\omega^{(n+1)}}{\partial T} 
    \bigg] =0\,, \\[2mm]
    \label{eq:expansions-2}
  &\Big(\frac{\partial \omega(U)}{\partial U}\Big)^\prime_\mathrm{S}  
   - \frac{\partial \omega(U)}{\partial U}  
   +\sum_{n=1}^\infty \lambda^n \bigg[ \frac1{\Delta_\mathrm{S}{}^{2n}}
    \,\Big(\frac{\partial \omega^{(n+1)}}{\partial
      U}
    \Big)^\prime_\mathrm{S} -   \frac{\partial
      \omega^{(n+1)}}{\partial U} \bigg] 
    =0\,,  \\[2mm]
    \label{eq:expansions-3}
     & 2\gamma \,\frac{\partial\log \Delta_\mathrm{S}} {\partial S} + 
      \frac1{\Delta_\mathrm{S}{}^2} \,
      \Big(\frac{\partial\omega(S)}{\partial S}\Big)^\prime_\mathrm{S}  
   - \frac{\partial \omega(S)}{\partial S}    \\
   &\quad +\sum_{n=1}^\infty \lambda^n \bigg[ \frac1{\Delta_\mathrm{S}{}^{2n+2}}
    \,\Big(\frac{\partial \omega^{(n+1)}}{\partial S}
    \Big)^\prime_\mathrm{S} -   \frac{\partial
      \omega^{(n+1)}}{\partial S} 
    - n\, \frac{\partial\log\Delta_{\mathrm{S}}}{\partial S} 
    \Big[\frac1{\Delta_\mathrm{S}{}^{2n}} \big(\omega^{(n+1)}
    \big)^\prime_\mathrm{S} 
  +\omega^{(n+1)} \Big] \bigg] 
    =0\,. \nonumber 
\end{align}
The above equations should hold for arbitrary values of $\lambda$.
Furthermore we remind the reader that the expressions with a prime
attached depend on the transformed fields $S'$, $T'$ and $U'$.  Upon
Taylor expanding in powers of $T'-T$ and $U'-U$, one will generate
additional terms proportional to powers of $\lambda$. The definition
of $S'$ under S-duality does not involve the parameter $\lambda$ and
can be effected directly (for instance, by using equations such as
\eqref{eq:transf-higher-der-omega}).  The last equation of
\eqref{eq:S-invariance-STU} is more complicated as it involves a
double sum,
\begin{align}
  \label{eq:fourth-expansion}
   & \sum_{n=1}^\infty \, n\, \lambda^n \bigg[ \frac1{\Delta_\mathrm{S}{}^{2n}}
    \, \big(\omega^{(n+1)}\big)^\prime_\mathrm{S}  - \omega^{(n+1)}
    \bigg] 
    +2\,\frac{\partial \log\Delta_\mathrm{S}}{\partial S} \,\lambda  \nonumber\\
     &\quad\times \bigg[ \frac{\partial
    \omega(T)}{\partial T} + \sum_{p=1}^\infty \lambda^{p} \,
  \frac{\partial\omega^{(p+1)}(S,T,U)}{\partial T} \bigg]\,
  \bigg[ \frac{\partial
    \omega(U)}{\partial U} + \sum_{q=1}^\infty \lambda^{q} \,
  \frac{\partial\omega^{(q+1)}(S,T,U)}{\partial U} \bigg] =0 \,.
\end{align}
Note that there are additional equations associated with T- and
U-duality. Those follow immediately by applying triality to the ones
specified above.

We note that the terms of order $\lambda^0$ cancel by virtue of the
first equation \eqref{eq:transf-higher-der-omega}. Furthermore, at
order $\lambda$, one directly derives the form of $\omega^{(2)}$ from
\eqref{eq:fourth-expansion},  
\begin{align}
  \label{eq:omega-2}
  &\omega^{(2)} (S,T,U) = \frac1{\gamma} \, \frac{\partial
    \omega}{\partial S}\,\frac{\partial \omega}{\partial
    T}\,\frac{\partial \omega}{\partial U}\,.
\end{align}
This result is consistent with triality and it satisfies all the other
equations
\eqref{eq:expansions-1}-\eqref{eq:expansions-3}.\footnote{
  The expression of $\omega^{(2)} (S,T,U)$ is unique. We have verified
  that it is not possible to add to it a covariant function
  $g(S, T, U)$ that is also triality invariant. Such a function would
  lead to a modification of $\omega^{(3)} (S,T,U)$ which would be
  incompatible with the transformation laws \eqref{eq:expansions-1}
  -\eqref{eq:expansions-3} and \eqref{eq:fourth-expansion}.  We expect
  that a similar finding applies to all the higher
  $\omega^{(n+1)} (S,T,U)$ with $n \geq 2$.}

Before continuing let us first note the systematics of the results
that gradually appears when working out all the variations. The power
of $\gamma$ that appear in the various terms of $\omega^{(n)}$ must be
equal to $n-k$, where $k$ is the number of functions $\omega$ that are
present. Furthermore every contributions must contain precisely $n-1$
derivatives with respect to $S$, $n-1$ with respect to $T$ and $n-1$
with respect to $U$. Finally $\omega^{(n)}$ will be multiplied by
$\lambda^{n-1}$ in the expansion \eqref{eq:Ups-holo-expansion}. This
pattern will persist in all the higher-order terms, something that can
be deduced from analyzing the original equations.

Let us now turn to the determination of the function
$\omega^{(3)}$, starting again with equation
\eqref{eq:fourth-expansion} and collecting all terms proportional to
$\lambda^2$. This requires to express the term $\omega^{(2)}(S',T',U')$ to
first order in $\lambda$, which yields
\begin{align}
  \label{eq:omega-2-prime}
  \frac1{\Delta_\mathrm{S}{}^2}\,  \omega^{(2)}(S',T',U') \to \,& 
  \frac{2\,\lambda}{\gamma}
  \,\frac{\partial\Delta_\mathrm{S}}{\partial S} 
  \Big(\frac{\partial\omega}{\partial S}  -2\gamma\,
    \frac{\partial\Delta_\mathrm{S}}{\partial S} \Big) \bigg[
    \frac{\partial^2 \omega}{\partial T^2}
    \,\Big(\frac{\partial\omega}{\partial U}\Big)^2  + 
    \frac{\partial^2 \omega}{\partial U^2}
    \,\Big(\frac{\partial\omega}{\partial T}\Big)^2\bigg] \,. 
\end{align}
Upon inspecting all possible terms contributing to
${\Delta_\mathrm{S}{}^{-2}} \, (\omega^{(3)})^\prime_\mathrm{S} -
\omega^{(3)}$, one easily verifies that the second derivatives
$\partial^2\omega/\partial T^2$ and $\partial^2\omega/\partial U^2$
appear, but there is no corresponding variation proportional to
$\partial^2\omega/\partial S^2$. This does not imply that triality
will be violated, simply because ${\Delta_\mathrm{S}{}^{-2}} \,
(\omega^{(3)})^\prime_\mathrm{S} - \omega^{(3)}$ will vanish for terms
that are proportional to the covariant combination $I^{(2)}(S)$
defined in \eqref{eq:def-I}. Hence one can include a term
$I^{(2)}(S)\,f(T,U)$ into $\omega^{(3)}$, where $f(T,U)$ can be chosen
such that the result for $\omega^{(3)}$ becomes consistent with
triality. In this way one derives the result
\begin{align}
  \label{eq:omega-3-1}
  \omega^{(3)} (S,T,U) =\;& -2\,
  \frac{\partial^2\omega}{\partial S^2}\, 
  \frac{\partial^2\omega}{\partial T^2}\,
  \frac{\partial^2\omega}{\partial U^2} \\
  &\;   -\frac1{\gamma} \bigg[ 
   \Big(\frac{\partial \omega}{\partial S}\Big)^2\,
    \frac{\partial^2\omega}{\partial T^2}\,
  \frac{\partial^2\omega}{\partial U^2} + 
   \frac{\partial^2\omega}{\partial S^2}\, 
   \Big(\frac{\partial \omega}{\partial T}\Big)^2\,
  \frac{\partial^2\omega}{\partial U^2}  +
  \frac{\partial^2\omega}{\partial S^2}\, 
  \frac{\partial^2\omega}{\partial T^2}\,
  \Big(\frac{\partial \omega}{\partial U}\Big)^2  \bigg]\nonumber\\
  &\; 
  + a_3\, \bigg[\frac{\partial^2\omega}{\partial S^2} +
  \frac1{2\gamma} \,
  \Big(\frac{\partial\omega}{\partial S}\Big)^2\bigg] \,
  \bigg[\frac{\partial^2\omega}{\partial T^2} + \frac1{2\gamma} \,
  \Big(\frac{\partial\omega}{\partial T}\Big)^2\bigg]\,
  \bigg[\frac{\partial^2\omega}{\partial U^2} +\frac1{2\gamma} \,
  \Big(\frac{\partial\omega}{\partial U}\Big)^2\bigg]  \,. \nonumber
\end{align}
Observe that we have included also an STU-covariant term that is invariant under triality with an
undetermined coefficient $a_3$. As it turns out this coefficient will
only be determined at the next order by requiring that $\omega^{(4)}$
will be triality invariant. We stress once more that derivatives of the function
$\omega$ can only depend on a single variable, $S$, $T$ or $U$.
The same conclusion holds for the other equations
\eqref{eq:expansions-1}-\eqref{eq:expansions-3} where the extra term
proportional to  $a_3$ does not contribute either. For the first two
equations this is obvious and for the third one one must make use of
the fact that 
\begin{equation}
  \label{eq:de-S-I-2}
 \Big(\frac{\partial{I^{(2)}(S)}} {\partial S}\Big)'_\mathrm{S} = 
  \Delta_\mathrm{S}{}^6 \,
  \bigg(\frac{\partial{I^{(2)}(S)}}{\partial S}  +4\,\frac{\partial\log
    \Delta_\mathrm{S}}{\partial S}  \, {I^{(2)}(S)} \bigg) \,. 
\end{equation}

Let us now continue the analysis to order $\lambda^3$ and consider
$\omega^{(4)}$. Following the same steps we find the following
expression for $\omega^{(4)}$ that is required by S-duality, 
without insisting on triality,
\begin{align}
  \label{eq:omega-4}
  \omega^{(4)} =&\; -\frac2{\gamma} \,\frac{\partial\omega}{\partial
    S} \,  \frac{\partial^2\omega}{\partial S^2}  \bigg[
  \frac{\partial^3\omega}{\partial T^3}\,
  \frac{\partial^2\omega}{\partial U^2}\,
  \frac{\partial\omega}{\partial  U} 
  +\frac1{2\gamma} \, \frac{\partial^3\omega}{\partial T^3}\, 
  \Big(\frac{\partial\omega}{\partial  U}\Big)^3 + 
  \frac1{\gamma} \,\frac{\partial^2\omega}{\partial T^2}\,
  \frac{\partial\omega}{\partial  T} \,  \frac{\partial^2\omega}{\partial U^2}\,
  \frac{\partial\omega}{\partial  U}  +T\leftrightarrow U \bigg]
  \nonumber\\[1mm]
  &\; -\frac1{\gamma^2} \bigg(\frac{\partial\omega}{\partial  S}\Big)^3
  \bigg[ \frac{\partial^3\omega}{\partial T^3}\,
  \frac{\partial^2\omega}{\partial U^2}\, 
  \frac{\partial\omega}{\partial  U}  +    \frac1{3} 
  \frac{\partial^3\omega}{\partial T^3}\,
  \Big(\frac{\partial\omega}{\partial  U}\Big)^3
  +T\leftrightarrow U  \bigg]
  \nonumber\\[1mm] 
  &\; -\frac1{\gamma^3} \Big(\frac{\partial\omega}{\partial  S}\Big)^3
  \, \frac{\partial^2\omega}{\partial T^2}\,
  \frac{\partial\omega}{\partial  T} \,\frac{\partial^2\omega}{\partial U^2}\,
  \frac{\partial\omega}{\partial  U}   \nonumber\\[1mm]
  &\;+ \frac{a_3}{\gamma}  \bigg[\frac{\partial^2\omega}{\partial S^2}
  \,\frac{\partial\omega}{\partial S}+
  \frac1{2\gamma} \,
  \Big(\frac{\partial\omega}{\partial S}\Big)^3\bigg] \,
  \bigg[\frac{\partial^2\omega}{\partial T^2} \,
  \frac{\partial\omega}{\partial T}+ \frac1{2\gamma} \,
  \Big(\frac{\partial\omega}{\partial T}\Big)^3\bigg]\,
  \bigg[\frac{\partial^3\omega}{\partial U^3}
  +\frac1{\gamma} \,
  \Big(\frac{\partial\omega}{\partial U}\Big)^3\bigg] \nonumber\\[1mm]
  &\; +  \frac{a_3}{\gamma} \bigg[\frac{\partial^2\omega}{\partial
    S^2} \,\frac{\partial\omega}{\partial S}+ \frac1{2\gamma} \,
    \Big(\frac{\partial\omega}{\partial S}\Big)^3\bigg]
    \,\bigg[\frac{\partial^3\omega}{\partial
      T^3}   + \frac1{\gamma} \,
    \Big(\frac{\partial\omega}{\partial T}\Big)^3\bigg]\,
     \bigg[\frac{\partial^2\omega}{\partial U^2} \,
  \frac{\partial\omega}{\partial U}+ \frac1{2\gamma} \,
  \Big(\frac{\partial\omega}{\partial U}\Big)^3\bigg] \,.
\end{align}
When insisting on triality it turns out that one must choose
$a_3=2$. For this value of $a_3$ it turns out that there is a
remarkable number of cancellations in $\omega^{(3)}$, whose final
expression takes the form
\begin{align}
  \label{eq:higher-orders-3}
  \omega^{(3)} =&\, \frac{1}{4\gamma^3} \, 
  \Big(\frac{\partial\omega}{\partial S}  \,
  \frac{\partial\omega}{\partial T}  \,
  \frac{\partial\omega}{\partial U}\Big)^2 \nonumber\\
  &\; + \frac1{2\gamma^2} \,\bigg[
  \frac{\partial^2\omega}{\partial S^2}
  \,\Big(\frac{\partial\omega}{\partial
    T}\,\frac{\partial\omega}{\partial U}\Big)^2 
  + \frac{\partial^2\omega}{\partial T^2}\,
  \Big(\frac{\partial\omega}{\partial U}\,
  \frac{\partial\omega}{\partial S} \Big)^2 
  + \frac{\partial^2\omega}{\partial U^2}
  \,\Big(\frac{\partial\omega}{\partial S}\,
  \frac{\partial\omega}{\partial T}  \Big)^2\bigg] \,. 
\end{align}
One then obtains the following result for $\omega^{(4)}$, 
\begin{align}
  \label{eq:higher-order-4}
  \omega^{(4)} =&\;  \frac1{6\,\gamma^5}   
  \Big(\frac{\partial\omega}{\partial S}  \,
  \frac{\partial\omega}{\partial T}  \,
  \frac{\partial\omega}{\partial U}\Big)^3    \nonumber\\
  &\; + \frac1{2\,\gamma^4} \bigg[   
    \frac{\partial^2\omega}{\partial S^2 }\,
    \frac{\partial\omega}{\partial S} \,\Big( 
    \frac{\partial\omega}{\partial T}\, 
    \frac{\partial\omega}{\partial U}\Big)^3 + 
    \frac{\partial^2\omega}{\partial T^2 }\,
    \frac{\partial\omega}{\partial T} \,\Big(
    \frac{\partial\omega}{\partial U}\,
    \frac{\partial\omega}{\partial S}\Big)^3
    +\frac{\partial^2\omega}{\partial U^2 }\,
    \frac{\partial\omega}{\partial U} \,\Big(
    \frac{\partial\omega}{\partial S}\,
    \frac{\partial\omega}{\partial T}\Big)^3 \bigg]  \nonumber\\   
&\; + \frac1{\gamma^3} \,\bigg[\Big(\frac{\partial\omega}{\partial S}\Big)^3\,
   \frac{\partial^2\omega}{\partial
    T^2 }\,\frac{\partial\omega}{\partial T}\,
  \frac{\partial^2\omega}{\partial U^2}\,
  \frac{\partial\omega}{\partial U} +
  \Big(\frac{\partial\omega}{\partial T}\Big)^3\,
   \frac{\partial^2\omega}{\partial
    U^2 }\,\frac{\partial\omega}{\partial U}\,
  \frac{\partial^2\omega}{\partial S^2}\,
  \frac{\partial\omega}{\partial S} \nonumber\\
  &\,\qquad\quad \quad +
  \Big(\frac{\partial\omega}{\partial U}\Big)^3\,
   \frac{\partial^2\omega}{\partial
    S^2 }\,\frac{\partial\omega}{\partial S}\,
  \frac{\partial^2\omega}{\partial T^2}\,
  \frac{\partial\omega}{\partial T}\bigg] \nonumber\\ 
  &\; +\frac1{6\gamma^3} \bigg[
  \frac{\partial^3\omega}{\partial S^3} 
  \,\Big(\frac{\partial\omega}{\partial T}\, 
  \frac{\partial\omega}{\partial U}\Big)^3 + 
  \frac{\partial^3\omega}{\partial T^3} 
  \,\Big(\frac{\partial\omega}{\partial U}\, 
  \frac{\partial\omega}{\partial S}\Big)^3 
  + \frac{\partial^3\omega}{\partial U^3} 
  \,\Big(\frac{\partial\omega}{\partial S}\, 
  \frac{\partial\omega}{\partial T}\Big)^3 \bigg]    \nonumber\\
  &\; + a_4\,\gamma\,\bigg[ \frac{\partial^3\omega}{\partial S^3} +
  \frac{3}{\gamma} \,
    \frac{\partial^2\omega}{\partial S^2}
    \,\frac{\partial\omega}{\partial S} + \frac1{\gamma^2} 
  \Big(\frac{\partial\omega}{\partial S}\Big)^3\bigg] \,
  \bigg[ \frac{\partial^3\omega}{\partial T^3} +
  \frac{3}{\gamma} \,
    \frac{\partial^2\omega}{\partial T^2}
    \,\frac{\partial\omega}{\partial T} + \frac1{\gamma^2} 
 \Big(\frac{\partial\omega}{\partial T}\Big)^3\bigg] \nonumber\\
  &\; \qquad\quad\quad   \times\bigg[ \frac{\partial^3\omega}{\partial U^3} +
  \frac{3}{\gamma} \,
    \frac{\partial^2\omega}{\partial U^2}
    \,\frac{\partial\omega}{\partial U} + \frac1{\gamma^2}  
    \Big(\frac{\partial\omega}{\partial U}\Big)^3\bigg] \,,
\end{align}
where, again, we introduced a new STU-covariant term proportional  to
the parameter $a_4$, which is triality invariant.  The value of $a_4$
is again expected to be fixed by insisting on triality in the next
order. 

Finally we consider the terms proportional to $\lambda^4$ and
concentrate on the solution for $\omega^{(5)}$. Based on S-duality
alone and using the result for the $\omega^{(n)}$ with $n<5$, we
arrange the result into an expression symmetric under triality, and
two sets of remaining terms. The triality symmetric expression, which
will constitute the final result, reads as follows (we organise the
terms in inverse powers of $\gamma$),
  \begin{align}
  \label{eq:omega-5}
  \omega^{(5)} =&\; \frac1{2\gamma^4} \bigg[
  \frac{\partial^3\omega}{\partial S^3}
  \frac{\partial{\omega}}{\partial{S}} \, \Big[
  \Big(\frac{\partial{\omega}}{\partial{T}}\Big)^2 \,
  \frac{\partial^2{\omega}}{\partial{T^2}}
  \,\Big(\frac{\partial{\omega}}{\partial{U}}\Big)^4 +
  \Big(\frac{\partial{\omega}}{\partial{U}}\Big)^2 \,
  \frac{\partial^2{\omega}}{\partial{U^2}}
  \,\Big(\frac{\partial{\omega}}{\partial{T}}\Big)^4 \Big]\nonumber\\
  &\; \quad\quad + \frac{\partial^3\omega}{\partial T^3}
  \frac{\partial{\omega}}{\partial{T}} \, \Big[
  \Big(\frac{\partial{\omega}}{\partial{U}}\Big)^2 \,
  \frac{\partial^2{\omega}}{\partial{U^2}}
  \,\Big(\frac{\partial{\omega}}{\partial{S}}\Big)^4 +
  \Big(\frac{\partial{\omega}}{\partial{S}}\Big)^2 \,
  \frac{\partial^2{\omega}}{\partial{S^2}}
  \,\Big(\frac{\partial{\omega}}{\partial{U}}\Big)^4
  \Big] \nonumber\\
  &\;\quad \quad + \frac{\partial^3\omega}{\partial U^3}
  \frac{\partial{\omega}}{\partial{U}} \, \Big[
  \Big(\frac{\partial{\omega}}{\partial{S}}\Big)^2 \,
  \frac{\partial^2{\omega}}{\partial{S^2}}
  \,\Big(\frac{\partial{\omega}}{\partial{T}}\Big)^4 +
  \Big(\frac{\partial{\omega}}{\partial{T}}\Big)^2 \,
  \frac{\partial^2{\omega}}{\partial{T^2}}
  \,\Big(\frac{\partial{\omega}}{\partial{S}}\Big)^4 \Big]
  \bigg]  \nonumber\\[2mm]
   &\; + \frac4{\gamma^4} \, \frac{\partial^2\omega}{\partial S^2} \, 
  \Big(\frac{\partial\omega}{\partial S}\Big)^2 \,
  \frac{\partial^2\omega}{\partial T^2} \,
  \Big(\frac{\partial\omega}{\partial T}\Big)^2 \,
  \frac{\partial^2\omega}{\partial U^2} \, 
  \Big(\frac{\partial\omega}{\partial U}\Big)^2\nonumber\\
  &\; + \frac1{24\, \gamma^4} \,\bigg[ 
  \frac{\partial^4\omega}{\partial S^4}
  \Big(\frac{\partial\omega}{\partial T}\Big)^4 
   \Big(\frac{\partial\omega}{\partial U}\Big)^4 +
   \frac{\partial^4\omega}{\partial T^4}
   \Big(\frac{\partial\omega}{\partial U}\Big)^4 
   \Big(\frac{\partial\omega}{\partial S}\Big)^4 +
   \frac{\partial^4\omega}{\partial U^4}
   \Big(\frac{\partial\omega}{\partial S}\Big)^4 
   \Big(\frac{\partial\omega}{\partial T}\Big)^4 \bigg]
  \nonumber\\[1mm]
  &\; + \frac1{2\gamma^4} \bigg[
  \Big(\frac{\partial\omega}{\partial S}\Big)^4  
  \Big[   \Big(\frac{\partial^2\omega}{\partial T^2}\Big)^2 \,
  \frac{\partial^2\omega}{\partial U^2} \,
  \Big(\frac{\partial\omega}{\partial U}\Big)^2 +  
  \Big(\frac{\partial^2\omega}{\partial U^2} \Big)^2 \,
  \frac{\partial^2\omega}{\partial T^2} \, 
  \Big(\frac{\partial\omega}{\partial T}\Big)^2 \Big] \nonumber\\
   &\; \qquad \quad + \Big(\frac{\partial\omega}{\partial T}\Big)^4  \Big[
  \Big( \frac{\partial^2\omega}{\partial U^2} \Big)^2 \,
  \frac{\partial^2\omega}{\partial S^2} \,
  \Big(\frac{\partial\omega}{\partial S}\Big)^2 +  \Big(
  \frac{\partial^2\omega}{\partial S^2} \Big)^2 \,
  \frac{\partial^2\omega}{\partial U^2} \, 
  \Big(\frac{\partial\omega}{\partial U}\Big)^2 \Big] \nonumber\\
  &\; \qquad \quad + \Big(\frac{\partial\omega}{\partial U}\Big)^4  
  \Big[\Big(\frac{\partial^2\omega}{\partial S^2} \Big)^2 \, 
  \frac{\partial^2\omega}{\partial T^2} \,
  \Big(\frac{\partial\omega}{\partial T}\Big)^2 
  +  \Big( \frac{\partial^2\omega}{\partial T^2} \Big)^2 \, 
  \frac{\partial^2\omega}{\partial S^2} \,
  \Big(\frac{\partial\omega}{\partial S}\Big)^2 \Big] \bigg] 
  \nonumber\\[1mm]
  &\; + \frac1{4\gamma^5} \bigg[ 
  \Big(\frac{\partial\omega}{\partial S}\Big)^4  
  \Big(\frac{\partial\omega}{\partial T}\Big)^4 \,\Big[
  \Big(\frac{\partial^2\omega}{\partial U^2} \Big)^2 +
  \frac{\partial^3\omega}{\partial U^3} \, \frac{\partial
    \omega}{\partial U} \Big] +  
  \Big(\frac{\partial\omega}{\partial T}\Big)^4  
  \Big(\frac{\partial\omega}{\partial U}\Big)^4 \,\Big[
  \Big(\frac{\partial^2\omega}{\partial S^2} \Big)^2  +
  \frac{\partial^3\omega}{\partial S^3} \, \frac{\partial
    \omega}{\partial S}\Big] \nonumber\\
  &\; \qquad\quad + 
  \Big(\frac{\partial\omega}{\partial U}\Big)^4 
  \Big(\frac{\partial\omega}{\partial S}\Big)^4 \,\Big[
  \Big(\frac{\partial^2\omega}{\partial T^2} \Big)^2 + 
  \frac{\partial^3\omega}{\partial T^3} \, \frac{\partial
    \omega}{\partial T} \Big]\bigg]
  \nonumber\\[1mm]  
    &\; + \frac2{\gamma^5} \bigg[  \frac{\partial^2\omega}{\partial
      S^2}  \Big( \frac{\partial \omega}{\partial S} \Big)^2 \, 
    \Big( \frac{\partial \omega}{\partial T} \Big)^4 \,  
    \frac{\partial^2\omega}{\partial U^2}  
    \Big( \frac{\partial \omega}{\partial U} \Big)^2
    +  \frac{\partial^2\omega}{\partial T^2}  
    \Big( \frac{\partial \omega}{\partial T} \Big)^2 \,
    \Big( \frac{\partial \omega}{\partial U} \Big)^4 \,  
    \frac{\partial^2\omega}{\partial S^2}  
    \Big( \frac{\partial \omega}{\partial S} \Big)^2 \nonumber\\
    &\;  \qquad \quad 
    +  \frac{\partial^2\omega}{\partial U^2}  \Big( \frac{\partial
      \omega}{\partial U} \Big)^2 \, 
    \Big( \frac{\partial \omega}{\partial S} \Big)^4 \,  
    \frac{\partial^2\omega}{\partial T^2}  
    \Big( \frac{\partial \omega}{\partial T} \Big)^2 \bigg]
    \nonumber\\[1mm]
   &\; + \frac5{8\gamma^6} \bigg[ 
   \frac{\partial^2\omega}{\partial S^2}  
   \Big( \frac{\partial \omega}{\partial S} \Big)^2  
   \Big( \frac{\partial \omega}{\partial T} \Big)^4 
   \Big(\frac{\partial \omega}{\partial U} \Big)^4 +  
   \frac{\partial^2\omega}{\partial T^2}  
   \Big( \frac{\partial\omega}{\partial T} \Big)^2  
   \Big( \frac{\partial \omega}{\partial U} \Big)^4  
   \Big( \frac{\partial \omega}{\partial S} \Big)^4 \nonumber\\
   &\:\quad\qquad  + 
   \frac{\partial^2\omega}{\partial U^2}  
   \Big( \frac{\partial\omega}{\partial U} \Big)^2  
   \Big( \frac{\partial \omega}{\partial S} \Big)^4  
   \Big( \frac{\partial \omega}{\partial T} \Big)^4 \bigg]
   \nonumber\\[1mm]
 &\; + \frac5{32\,\gamma^7} \, \Big( \frac{\partial \omega}{\partial
   S} \Big)^4  \Big( \frac{\partial \omega}{\partial T} \Big)^4 \Big(
 \frac{\partial \omega}{\partial U} \Big)^4 \;. 
\end{align}
In addition there are two more contributions. One takes the form 
\begin{align}
  \label{eq:remainder-1}
  \big[\omega^{(5)}\big]_1 = &\; 
  - \frac1{2 \gamma^4} \, \big( I^{(2)} (S) \big)^2 
 \bigg[ \frac{\partial^2\omega}{\partial T^2} \,
  \Big(\frac{\partial\omega}{\partial T}\Big)^2
  \Big(\frac{\partial\omega}{\partial U}\Big)^4 +
  \frac{\partial^2\omega}{\partial U^2} \,
  \Big(\frac{\partial\omega}{\partial U}\Big)^2
  \Big(\frac{\partial\omega}{\partial T}\Big)^4 \bigg]
  \nonumber\\
   &\; -  \frac1{4\gamma^5}   \big( I^{(2)} (S) \big)^2  \Big(
   \frac{\partial \omega}{\partial T} \Big)^4  \Big( \frac{\partial
     \omega}{\partial U} \Big)^4  
   - \frac1{24 \, \gamma^4} \, I^{(4)} (S)
   \Big(\frac{\partial\omega}{\partial T}\Big)^4
   \Big(\frac{\partial\omega}{\partial U}\Big)^4  \,,
\end{align}
and is proportional to the S-covariant terms $[ I^{(2)} (S) ]^2$ and
$I^{(4)} (S)$. 
These terms 
are not fixed by the
equation for $\omega^{(5)}$.
The other contribution contains 
terms that
are proportional to the undetermined constant
$a_4$ introduced in \eqref{eq:higher-order-4}.
 They take the form
\begin{align}
  \label{eq:remainder-2}
  \big[\omega^{(5)}\big]_2 = &\; \frac{a_4}{\gamma} \bigg[
  \frac{\partial^3\omega}{\partial S^3} + \frac{3}{\gamma} \,
  \frac{\partial^2\omega}{\partial S^2}
  \,\frac{\partial\omega}{\partial S} + \frac1{\gamma^2}
  \Big(\frac{\partial\omega}{\partial S}\Big)^3\bigg] \,
  \frac{\partial{\omega}}{\partial{S}} \nonumber\\
  &\, \qquad \times \bigg[ \frac{\partial^4\omega}{\partial T^4} +
  \frac{3}{\gamma} \,\frac{\partial^3\omega}{\partial T^3}
  \,\frac{\partial\omega}{\partial T} + \frac{3}{\gamma} \,
  \Big(\frac{\partial^2\omega}{\partial T^2}\Big)^2 + \frac3{\gamma^2}
  \,\frac{\partial^2\omega}{\partial T^2}\,
  \Big(\frac{\partial\omega}{\partial T}\Big)^2\bigg] \nonumber\\
  &\; \qquad\times\bigg[ \frac{\partial^3\omega}{\partial U^3} +
  \frac{3}{\gamma} \, \frac{\partial^2\omega}{\partial U^2}
  \,\frac{\partial\omega}{\partial U} + \frac1{\gamma^2}
  \Big(\frac{\partial\omega}{\partial U}\Big)^3\bigg] \,
  \frac{\partial\omega}{\partial U} + [T\leftrightarrow U] \,,
\end{align}
This expression is not triality invariant, and neither can it be
made invariant by including terms of the form $I^{(4)}(S) \,f(T,U) +
\big[I^{(2)}(S)\big]^2 \,g(T,U)$. Hence it follows that $a_4=0$.
 
Therefore $\omega^{(5)}$ is given by \eqref{eq:omega-5}, up to
STU-covariant terms consisting of triality symmetric products of
$I^{(4)}(S)$ or $\big[I^{(2)}(S)\big]^2$, $I^{(4)}(T)$ or
$\big[I^{(2)}(T)\big]^2$, and $I^{(4)}(U)$ or
$\big[I^{(2)}(U)\big]^2$.  There are precisely four such terms,
multiplied by arbitrary constants and integer powers of $\gamma$
ranging between $\gamma^{-1}$ and $\gamma^2$. Based on the experience
for $n\leq4$ so far, we expect that these undetermined terms will be
fixed by proceeding with the present analysis to level
$\lambda^5$. However, in the next section we will change strategy, and
at the end of section \ref{sec:addit-contr-h1} we will then discover
an alternative way of proving that $\omega^{(5)}$ will precisely be
given by equation \eqref{eq:omega-5}.

Hence at this point we have determined the coefficient functions
$\omega^{(n)}$ for $n\leq 5$. We are not aware of any impediment when
continuing the present calculation to higher orders and expect that
the function $\Omega$ can be uniquely determined from STU-duality
combined with triality.

\section{The topological string partition function}
\label{sec:topol-string-part}
\setcounter{equation}{0}
It is possible to obtain a corresponding version of the topological
partition functions from the function \eqref{eq:holo-function} that
encodes the Wilsonian effective action. This relation involves a
Legendre transform and as a result the topological string will behave
differently under the duality symmetries. The transformation rules of
its moduli will not be affected by the possible introduction of
deformations, such as those associated with $\Omega$. Upon performing
the Legendre transform, one obtains the so-called Hesse potential
\cite{LopesCardoso:2006ugz}, which decomposes into an infinite variety
of different functions. One of these functions exhibits the
characteristic features of the topological string partition function.
As a result the moduli of the topological string are not identical to
the moduli that appear in the Wilsonian action. Actually the same
phenomenon is encountered in field theory when considering the
Lagrangian and the Hamiltonian of a four-dimensional theory with
abelian vector gauge fields.  The dynamical variables that appear in
the Lagrangian are different from those that appear in the
Hamiltonian, and the consequences of electric-magnetic duality will be
realized in a different way.

The Hesse potential is a real function of the moduli. As we will see,
its moduli will transform covariantly under the dualities and the
transformation rules do not change because of the presence of possible
deformations.  Hence the Hesse potential will transform as a {\it
  function} under general duality transformations, which in the model
at hand constitute the group $\mathrm{Sp}(8;\mathbb{R})$, and it will
remain invariant under the subgroup thereof equal to
$\Gamma_0(2)_\mathrm{S}\times\Gamma_0(2)_\mathrm{T}
\times\Gamma_0(2)_\mathrm{U}$. As was demonstrated in
\cite{Cardoso:2014kwa}, the Hesse potential can be obtained from the
function $F(X,A)$ that encodes the Wilsonian action. To construct the
Hesse potential it is important that the deformation $\Omega$ is real,
whereas in \eqref{eq:holo-function} it we assumed to depend only on
the holomorphic moduli. To make $\Omega$ real we will simply add its
complex conjugate. This change is not problematic as the holomorphic
derivatives $F_I$ are not affected, so that the results of the previous
section will remain valid. When deriving the expression for the Hesse
potential, we will {\it for the moment} replace $\Omega$ by
$\Omega(X,A)+\bar\Omega(\bar X,\bar A)$.

The Legendre transformation is most easily understood by first
considering a conversion to real special geometry, where the real
fields $\big(\phi^I, \chi_J\big)$ transform under the dualities precisely as
the dual pair $(X^I, F_J(X,A))$, where $F_J(X,A)$ denotes the
derivatives of the function \eqref{eq:holo-function} with respect to
the $X^J$. Hence we consider the redefinitions, 
\begin{equation}
  \label{eq:X-real}
  \phi^I= 2\,\mathrm{Re} \,X^I  \,,\qquad  \chi_J= 2\, \mathrm{Re} \,F_J(X,A) \,.
\end{equation}
The Hesse potential is now obtained by a Legendre transform with
respect to the imaginary part of the $X^I$ \cite{LopesCardoso:2006ugz},
\begin{equation}
  \label{eq:Hesse-general}
  \mathcal{H}(\phi,\chi)  = 4 \,\big[ \mathrm{Im} \,F(X) +
  \Omega(X,A) + \bar\Omega(\bar X,\bar A)\big] + \mathrm{i}\,\chi_I
  (X^I-\bar X^I) \,, 
\end{equation}
where $F(X)$ is equal to the function \eqref{eq:class-F}. The
Hesse potential transforms as a {\it function} under generic
$\mathrm{Sp}(8;\mathbb{R})$ dualities and is left {\it invariant}
under the 
$\Gamma_0(2)_\mathrm{S}\times\Gamma_0(2)_\mathrm{T}
\times\Gamma_0(2)_\mathrm{U}$ subgroup.

The topological string partition function are conventionally written
in terms of complex moduli that transform covariantly under the
dualities. Therefore we carry out a conversion by subsequently following the
inverse procedure \eqref{eq:X-real}, but to different moduli
$\mathcal{X}^I$,
\begin{align}
  \label{eq:X-calX}
  2\,\mathrm{Re} \,X^I  =&\,\, \phi^I
                           = 2 \, \mathrm{Re}\,\mathcal{X}^I\,,\nonumber\\ 
  2\,\mathrm{Re} \,F_I(X,A)  =&\, \chi_I =
                         2\,  \mathrm{Re}\,\mathcal{F}_I(\mathcal{X}) \,,
\end{align}
where on the left-hand side we have the original fields $X^I$ and the
$X^I$- derivatives of the function \eqref{eq:holo-function}, and on the
right-hand side the new fields $\mathcal{X}^I$ and the
$\mathcal{X}^I$-derivatives of the classical function
$\mathcal{F}(\mathcal{X})$, equal to 
\begin{equation}
  \label{eq:classical-F}
  \mathcal{F}(\mathcal{X}) =
  -\,\frac{\mathcal{X}^1\mathcal{X}^2\mathcal{X}^3}{\mathcal{X}^0}
  \,. 
\end{equation}
This relation is motivated by the fact that both sides of these
equations transform consistently under the same duality
transformations. Obviously the $X^I$ and $\mathcal{X}^I$ will differ
by terms proportional to powers of $\Omega$. The details of this
construction have been described in \cite{Cardoso:2014kwa}.  The next
step is to express the original moduli $X^I$ in terms of the new ones,
$\mathcal{X}^I$; this can be done by iteration.  The results can then
be substituted into the expression \eqref{eq:Hesse-general}. Similar
evaluations have been carried out in \cite{Cardoso:2010gc,
  Cardoso:2014kwa} for various models. However, to make contact with
the topological string partition function it is important to realize
that the Hesse potential will decompose into an infinite number of
functions that are separately invariant under the action of the
duality subgroup that constitute an invariance of the model. The
general situation may be described as follows,
\begin{align}
  \label{eq:Hesse-decomp}
  \mathcal{H} =&\, \mathcal{H}^{(0)} + \mathcal{H}^{(1)} +
  \mathcal{H}^{(2)} + \big(\mathcal{H}^{(3)}_1 + {\mathcal{H}}^{(3)}_2
  + \mathrm{h.c.}\big) + \mathcal{H}^{(3)}_3 +
  \mathcal{H}^{(4)}_1  +\mathcal{H}^{(4)}_2 +\mathcal{H}^{(4)}_3 \nonumber\\
  &\, +\big(\mathcal{H}^{(4)}_4+ \mathcal{H}^{(4)}_5 +
  \mathcal{H}^{(4)}_6 + \mathcal{H}^{(4)}_7 + \mathcal{H}^{(4)}_8 +
  \mathcal{H}^{(4)}_9 +\mathrm{h.c.}\big) \ldots\,.
\end{align}
The leading terms of some of these functions are presented in
\cite{Cardoso:2014kwa}. For the STU-model of this paper, each of these
functions will be invariant under $[\Gamma_0(2)]^3$ dualities.

The first function, $\mathcal{H}^{(0)}(\mathcal{X},\bar{\mathcal{X}})$
in \eqref{eq:Hesse-decomp} is simply the Hesse potential associated
with the classical function \eqref{eq:classical-F}, which is real and
non-holomorphic,
\begin{equation}
  \label{eq:Hesse-0}
  \mathcal{H}^{(0)}(\mathcal{X},\bar{\mathcal{X}})  = - \mathrm{i}\,\big[
  \bar{\mathcal{X}}^I \mathcal{F}_{I} (\mathcal{X}) 
  -\mathcal{X}^I \bar{\mathcal{F}}_{I}
  (\bar{\mathcal{X}}) \big]\,.
\end{equation}
As shown in \eqref{eq:Hesse-decomp}, there are infinitely many
additional functions that emerge which all depend on the extension
$\Omega$.  For a general real function $\Omega$, $\mathcal{H}^{(1)}$
has been presented up to terms of order $\Omega^5$. However, as was
pointed out earlier, for the STU-model the function $\Omega$ is
actually harmonic, so it can be written as the sum of a holomorphic
function $\Omega$ and its complex conjugate. Therefore it suffices to
only give the terms proportional to $A$, so that $\Omega$ will depend
holomorphically on the modular forms $\omega$,
\begin{align}
  \label{eq:Hesse-1}
  \mathcal{H}^{(1)} =\,\big[&  4\,\Omega - 4\,N^{IJ}\,\Omega_I\Omega_J
      +8\, \Omega_{IJ} (N \Omega)^I (N \Omega)^J
     \nonumber\\ 
    &\, + \tfrac{8}3 \,\mathrm{i} \mathcal{F}_{IJK} (N \Omega)^I (N
      \Omega)^J (N \Omega)^K
      \nonumber\\
    &\, - \tfrac43 \mathrm{i} \, \left(\mathcal{F}_{IJKL} + 3 \mathrm{i}
      \mathcal{F}_{R(IJ} N^{RS} \mathcal{F}_{KL)S} \right) (N
      \Omega)^I (N \Omega)^J (N \Omega)^K (N \Omega)^L
       \nonumber\\
     &\, 
    - \tfrac{16}3\,  \Omega_{IJK} (N \Omega)^I (N \Omega)^J (N
      \Omega)^K   \nonumber\\
      &\ - 16 \mathrm{i} \, \mathcal{F}_{IJK} N^{KP} \,\Omega_{PQ}  (N
      \Omega)^I (N \Omega)^J (N \Omega)^Q 
           \nonumber\\
     & \, - 16\,  (N \Omega)^P  \, \Omega_{PQ} \, N^{QR}
     \Omega_{RK} \, (N \Omega)^K  
    + \mathcal{O}(\Omega^5)\, \big] + \text{h.c.}\;.   
\end{align} 
Here we have used the notation $(N \Omega)^I = N^{IJ} \Omega_J$,
$(N \bar\Omega)^I = N^{IJ} \Omega_{\bar J}$, with $N^{IJ}$ the
 inverse of $N_{IJ}=
 2\,\mathrm{Im}\,[\mathcal{F}(\mathcal{X})_{IJ}]$. Explicit
 expressions for both $N_{IJ}$ and $N^{IJ}$ are given in appendix
 \ref{sec:formulae}. Note that the whole expression is now written in
 terms of the new moduli $\mathcal{X}^I$, and no longer in terms of
 the original moduli. 

All other functions in \eqref{eq:Hesse-decomp} are qualitatively very
different from $\mathcal{H}^{(1)}(\mathcal{X},\bar{\mathcal{X}})$: they
do not contain terms linear in $\Omega$ and they are not harmonic in
$\Omega$. These special features identify the function
$\mathcal{H}^{(1)}$ as the unique candidate for the topological string
partition function.  However, it is important to appreciate that
$\mathcal{H}^{(1)}$ is {\it not} holomorphic in $\mathcal{X}^I$ in
view of the presence of the tensors $N^{IJ}$. The results of the previous
sections based on \eqref{eq:Ups-holo-expansion} imply that the
function $\mathcal{H}^{(1)}$ does instead have the following form,
\begin{align}
  \label{eq:H-1}
  \mathcal{H}^{(1)} = 4A\,\big[- \gamma\, \ln \lambda + h(\omega;
  \lambda)\,\big] + \text{h.c.}  \,,
\end{align}
where the function $h(\omega;\lambda)$ depends on the
holomorphic modular form $\omega$ and its covariant derivatives,
and on the holomorphic topological string coupling constant $\lambda$
defined below. While the modular form $\omega$ depends only on the
moduli $S,T,U$, we will see that the covariant derivatives involve a
non-holomorphic connection, so that the function $h(\omega;\lambda)$ 
will depend explicitly on $S$, $T$, $U$, and their complex
conjugates. This feature is characteristic for the topological string
partition function. 

Incidentally, we note that when suppressing all the non-holomorphic
terms in \eqref{eq:Hesse-1} the function $h(\omega;\lambda)$ will be
holomorphic and only the holomorphic $\Omega$ will remain. In that
limit there is no longer a distinction between the old and the new
moduli so that $h(\omega;\lambda)$ must become equal to
\begin{equation}
  \label{eq:hole-limit-h}
  h(\omega;\lambda)\longrightarrow \omega^{(1)}(S,T,U) +
  \sum_{n=1}^{\infty} \, 
  \lambda^{n} \, \omega^{(n+1)}(S,T,U)\,. 
\end{equation}
One can use this observation to relate specific terms in
\eqref{eq:Hesse-1} to the terms in the holomorphic function that was
evaluated in section \ref{sec:higher-orders}. We will make use of this
observation at the end of section \ref{sec:addit-contr-h1} and in
appendix \ref{sec:Large-moduli-limits}.

Let us now turn to the precise nature to the non-holomorphic
terms. First of all, the new moduli and the parameter $\lambda$ are
defined by
\begin{equation}
  \label{eq:def-STU-covariant}
  S=-\mathrm{i} \,\frac{\mathcal{X}^1}{\mathcal{X}^0} \,,\qquad
  T=-\mathrm{i}
  \,\frac{\mathcal{X}^2}{\mathcal{X}^0} \,,\qquad 
  U=-\mathrm{i} \,\frac{\mathcal{X}^3}{\mathcal{X}^0}\qquad \lambda=
  \frac{A}{(\mathcal{X}^0)^2}  \,,   
\end{equation}
which are similar but not identical to ones defined in previous
sections. The fields and $\lambda$ transform under S-duality as
\begin{equation}
  \label{eq:STU-S-dual-covariant}
  S\to \frac{a\,S-\mathrm{i} b}{d+ \mathrm{i}c\,S} \,,\qquad T\to
  T\,,\qquad U\to U\,, \qquad 
  \lambda \to  \frac{\lambda}{(d+ \mathrm{i}c\,S)^2}  \,,
\end{equation}
but one should keep in mind that these fields are fundamentally
different from the original ones defined in \eqref{eq:def-STU},
because their duality transformations \eqref{eq:STU-S-dual-covariant}
are exact and will not be affected by the presence of
$\Omega(X,A)$. As before, the corresponding transformations under T-
and U-duality follow from triality.

We explicitly evaluate the first few terms of the function
$h(\omega;\lambda)$ to appreciate how its duality invariance is
realized.  This follows from repeated use of the following identity,
where $V$ and $W$ are two arbitrary functions that depend
holomorphically on $\mathcal{X}^I$, 
\begin{align}
  \label{eq:partial-N-partial}
  \frac{\partial V}{\partial\mathcal{X}^I} \, N^{IJ} \, \frac{\partial
    W}{\partial{\mathcal{X}}^J} =&\,\frac1{(\mathcal{X}^0)^2}
  \,\sum_{STU} \frac1{S+\bar S} \, \frac{\partial V}{\partial T}
  \,\frac{\partial
    W}{\partial U} \nonumber\\
  &\, -\frac1{\mathcal{X}^0} \,\sum_{STU} \Big[\frac{\partial
    V}{\partial\mathcal{X}^0} \,
  \frac1{(S+\bar S)(T+\bar T)} \,  
  \,\frac{\partial W}{\partial U} +\big\{ V\leftrightarrow W \big\}
  \Big]\nonumber \\
  &\, + \frac2{(S+\bar S) (T+\bar T) (U+\bar U) }
  \,\frac{\partial V}{\partial\mathcal{X}^0} \,
  \,\frac{\partial W}{\partial\mathcal{X}^0}\,,
\end{align}
where we used the explicit expression for the matrix $N^{IJ}$ given in
\eqref{eq:N-IJ-inverse} and where $\sum_{STU}$ denotes the sum over all
independent permutations of $\{S,T,U\}$.

Using the explicit expressions in section \ref{sec:higher-orders}, one
then obtains the following result for $h(\omega; \lambda)$, 
\begin{align}
  \label{eq:topological-string}
   &h(\omega; \lambda) = 
  \omega(S) + \omega(T) + \omega(U) 
  + \frac{\lambda} {\gamma} \,\Big[ D_S\,\omega \; D_T\,\omega
  \;D_U\, \omega\Big]  \nonumber\\
  &\,\qquad+ \lambda^2 \,
  \bigg[ \frac1{4\gamma^3} \, (D_S\,\omega)^2 \,( D_T\,\omega)^2  \,
  (D_U\,\omega)^2 \nonumber\\ 
  &\qquad\quad\qquad
  +\frac1{2\gamma^2} \Big[ (D_S{}^2\omega) \, (D_T\,\omega)^2
  \,(D_U\,\omega)^2 + (D_S\,\omega)^2 \, (D_T{}^2\omega)
  \,(D_U\,\omega)^2 \nonumber\\
  &\qquad\qquad\qquad\qquad \qquad \qquad
  +  (D_S\,\omega)^2 \, (D_T\,\omega)^2
  \,(D_U{}^2\omega) \Big]\, \bigg]    \nonumber \\
 &\,\qquad + \lambda^3 \,
  \bigg[ \frac1{6\gamma^5} \, (D_S\,\omega)^3 \,( D_T\,\omega)^3  \,
  (D_U\,\omega)^3 \nonumber\\ 
   &\qquad\quad\qquad
  +\frac1{2\gamma^4} \Big[ (D_S{}^2\omega) \, (D_S\omega) \, (D_T\,\omega)^3
  \,(D_U\,\omega)^3 \nonumber\\   
   &\qquad\qquad \qquad\quad\quad 
   + (D_S\,\omega)^3 \, (D_T{}^2\omega)
  \, (D_T \, \omega) 
  \,(D_U\,\omega)^3 \nonumber\\[1mm]
  &\qquad\qquad\qquad\quad\quad
  +  (D_S\,\omega)^3 \, (D_T\,\omega)^3
  \,(D_U{}^2\omega) \, (D_U \, \omega) \Big]     \nonumber \\
    &\qquad\quad\qquad
  +\frac1{\gamma^3} \Big[ (D_S\,\omega)^3 \,  (D_T{}^2\omega) \, (D_T
  \, \omega)  
  \,   (D_U{}^2\omega) \, (D_U \, \omega) \nonumber\\
     &\qquad\qquad \qquad \qquad
     + (D_S{}^2\omega) \, (D_S \, \omega) \, (D_T{}^2\omega) \, (D_T
     \, \omega) 
     \, (D_U \, \omega)^3 \nonumber\\[1mm]
 &\qquad\qquad\qquad\qquad
 + (D_S{}^2\omega) \, (D_S \, \omega) \,  (D_T \, \omega)^3 \, (D_U{}^2\omega) \, (D_U \, \omega) \Big]
 \nonumber\\  
   &\qquad\quad\qquad
  +\frac1{6\gamma^3} \Big[ (D_S{}^3\omega) \,  (D_T\,\omega)^3 \,
  (D_U\,\omega)^3 
  +  (D_S\,\omega)^3 \, (D_T{}^3\omega)  \, (D_U\,\omega)^3 \nonumber\\
  &\qquad\qquad\qquad\qquad  +  (D_S\,\omega)^3 \, (D_T \omega)^3  \,
  (D_U{}^3\omega) \Big] \bigg] \nonumber\\
  &\,\qquad+ \mathcal{O}(\lambda^4)  \,,
\end{align}
where we have introduced non-holomorphic duality covariant derivatives
defined such that
\begin{equation}
D_S{}^n \omega \to  \Delta_{\mathrm S}{}^{2n} \, D_S{}^n \omega\,, 
\end{equation}
which shows that the expression \eqref{eq:H-1} is manifestly
STU-duality invariant as it should. Incidentally, the duality
transformation of the functions $\omega$ may involve a constant
imaginary shift due to the multiplyer system, as discussed in section
\ref{sec:STU-model-dualities}, which will cancel in the variation of
\eqref{eq:genus-1} below. Explicit expressions for the covariant derivatives
are, for instance,
\begin{align}
  \label{eq:D-omega}
  D_S \,\omega = \frac{\partial \omega}{\partial S} -\frac{
    2\gamma}{S + \bar S}\,, \qquad
  D_S{}^2 \omega = \frac{\partial^2 \omega}{\partial S^2} 
  +\frac{2}{(S + \bar S)}\,\frac{\partial\omega}{\partial S} 
  -\frac{2\gamma}{(S + \bar S)^2}\,,
\end{align}
while for a covariant quantity $\Sigma(S)$ of weight $p$, which
transforms under S-duality according to $\Sigma(S)\to
\Delta_\mathrm{S}(S){}^{p}\, \Sigma(S)$, the covariant derivative equals 
\begin{equation}
  \label{eq:D-Sima}
   D_S\,\Sigma(S) = \Big(\frac{\partial}{\partial S} +\frac{
    p }{S + \bar S}\Big)\,\Sigma(S) \,.
\end{equation}
These results can be combined with similar expressions that involve
the Serre derivative. For instance, we note the convenient identity
\begin{equation}  
  \label{eq:I-n+1-nonholo-der}
  I^{(n+1)}(S) = D_S I^{(n)} (S) + \frac{n}{\gamma} \, I^{(n)}(S)
  \,D_S \,\omega \,,
\end{equation}
which follows from \eqref{eq:DI-I}. Another useful identity is,
\begin{equation}
  \label{eq:D2-omega}
  D_S{}^2 \,\omega= I^{(2)}(S) - \frac1{2\gamma}\,
  \big(D_S\,\omega\big)^2 \,. 
\end{equation}

Let us now return to $h(\omega; \lambda)$ and rewrite it as follows,
\begin{equation}
  \label{eq:topological-string-g-exp}
   h(\omega; \lambda) = 
  \omega(S) + \omega(T) + \omega(U) 
  + \sum_{g=2}^\infty \, \lambda^{g-1} F^{(g)}(S,T,U) \,. 
\end{equation} 
This expression defines the genus expansion of the topological string
partition function with $g\geq2$, where the $F^{(g)}$ depend on the
functions $\omega(S)$, $\omega(T)$ and $\omega(U)$ and their covariant
derivatives, which depend on $S,T,U$ and their complex
conjugates. Here $\lambda$ plays the role of the topological string
coupling constant. As an example we present the expression for
$F^{(2)}$,
\begin{equation}
  \label{eq:genus2}
    F^{(2)}(S,T,U)  = \frac{\lambda}{\gamma} \,
    D_S\,\omega(S)\;D_T\,\omega(T) 
  \;D_U\,\omega(U) \,, 
\end{equation}
where we stress that the dependence on $\bar{S}$, $\bar{T}$ and
$\bar{U}$ is implicit and contained in the covariant derivatives. For
higher genus $g=3,4$ the result can be read off from
\eqref{eq:topological-string}. Based on \eqref{eq:Hesse-1}, we also
obtain the genus-1 partition function (which is real and harmonic),
\begin{equation}
  \label{eq:genus-1}
  F^{(1)}= -\gamma \,\ln\vert\lambda\vert^2 + \omega(S) + \omega(T) + \omega(U) 
  + \bar\omega(\bar S) + \bar\omega(\bar T) +
  \bar\omega(\bar U)  \,, 
\end{equation}
where here and henceforth we choose $A$ equal to unity.
  
For $g\geq 2$ the partition functions satisfy a holomorphic anomaly
equation of the form 
\begin{align}
  \label{eq:hol-anomaly-eq}
  \frac{\partial h}{\partial \bar S}= \frac{2\lambda}{(S+\bar S)^2} \,
  D_T\, h \,D_U\, h \,, 
\end{align}
which can be verified up to order $\lambda^4$ on the basis of the
results obtained so far, with similar equations for the $\bar{T}$ and $\bar{U}$
derivatives. Observe that the anti-holomorphic derivative with respect
to $\bar{\mathcal{X}}^0$ vanishes, as the dependence on
$\lambda$ is holomorphic. The equation \eqref{eq:hol-anomaly-eq} was 
encountered in \cite{Cardoso:2014kwa}  as a result of the diagrammatic
structure in the Hesse potential. The same arguments apply in this
case, so that we may assume that \eqref{eq:hol-anomaly-eq} holds
to all orders. 

Note that the above results do not entirely agree with the holomorphic
anomaly equation obtained in
\cite{Bershadsky:1993ta,Bershadsky:1993cx}, in particular because
$F^{(1)}$ is harmonic and therefore not affected by the
anomaly. However, one can replace the term
$-\gamma \ln\vert\lambda\vert^2$ by $-\gamma \ln N$, where
$N=\vert\det N_{IJ}\vert$, since they transform identically under
duality. In that case one obtains
\begin{equation}
  \label{eq:F-1-new}
  F^{(1)} = - \gamma \, \ln N + \omega(S)+ \omega(T)
+ \omega(U) + \bar\omega(\bar S) + \bar\omega(\bar T) +
\bar\omega(\bar U)\,. 
\end{equation}
Equivalently, one could take the view that we have introduced an extra
term equal to $-\gamma \big[ \ln N - \ln\vert\lambda\vert^2$ which is
duality invariant and non-harmonic. It seems obvious that this
modification will not affect the higher-order terms of
$h(\omega;\lambda)$, because those do already respect the duality
invariance.

On the other hand, in \cite{Cardoso:2014kwa} we have demonstrated how
a non-harmonic term in $F^{(1)}$, which transforms into harmonic
variations under duality, will introduce non-harmonic terms into the
higher-genus contributions. We have therefore explicitly verified that
\eqref{eq:F-1-new} will indeed induce the same non-holomorphic
corrections as we have found earlier in \eqref{eq:topological-string}.

\section{Partial determination of the function $h$}
\label{sec:partial-det-h}
\setcounter{equation}{0}
Here and in the next section we will try to further determine the
function $h(\omega;\lambda)$ that comprises the part of the topological
string partition function that depends holomorphically on the
topological string coupling constant $\lambda$.
Inspired by \cite{Alim:2015qma}, we therefore consider
the special limit where the modular forms $\omega$
are suppressed and consider the possibility of an exact expression for
the topological string partition function. The expression for the
function $h$ then takes the form of a power series in terms of an
effective coupling constant $\tilde\lambda$ defined by
\begin{equation}
  \label{eq:eff-lambda}
  \tilde \lambda= \frac{\lambda}{(S+\bar S)(T+\bar T)(U+\bar U)}\,.
\end{equation}
Indeed, suppressing $\omega$  in \eqref{eq:topological-string} leads to
\begin{align}
  \label{eq:exp-h}
  h_0(\tilde\lambda)=&\, \sum_{n=2} a_n\, \tilde\lambda^{n-1} 
  =-8\,\gamma^2\,\tilde \lambda -32\,\gamma^3
  \,\tilde\lambda^2  + \mathcal{O}(\tilde\lambda^3) \,, 
\end{align}
where we appended the subscript to indicate that this is only a
truncated version of the original function $h(\omega;\lambda)$. Here we
note that $\tilde\lambda$ transforms under duality with a phase,
e.g. $\tilde\lambda\to
\big(\bar{\Delta}_\mathrm{S}/{\Delta}_\mathrm{S}\big)\,\tilde\lambda$. However,
we will eventually replace $\tilde\lambda$ by a modified expansion
parameter that is fully STU invariant. The main topic of this section
is to determine the exact expression for $h_0$ and to derive an
equation for the additional terms which will be contained in another
function $h_1$. 

Subsequently we substitute \eqref{eq:exp-h} into the non-linear equation
\eqref{eq:hol-anomaly-eq}. Before doing so we first evaluate the
result for $D_T\,h$ and $\partial_{\bar S}\,h$ upon suppressing $\omega$,  
\begin{align}
  \label{eq:3}
  (T+\bar T)\,D_T\,h\Big\vert_{\omega=0} =&\, -2\gamma + \sum_{n=2}
  \,(n-1) \,a_n\,   \tilde\lambda^{n-1} \,,\nonumber\\
  (S+\bar S) \,\frac{\partial h}{\partial\bar S}\Big\vert_{\omega=0}=
  &\, - \sum_{n=2}\,   (n-1) \,a_n \,\tilde\lambda^{n-1} \,.
\end{align}
Equation \eqref{eq:hol-anomaly-eq} then leads to $a_2= -8\,\gamma^2$
and $a_3=-32\,\gamma^3$ by considering the terms proportional to
$\tilde\lambda$ and $\tilde\lambda^2$, respectively, which is in
agreement with the values found in \eqref{eq:exp-h}. The terms in
higher powers of $\tilde\lambda$ then yield the following
equations (for $n\geq4$),
\begin{equation}
  \label{eq:fixed-power}
  -(n-1) \,a_n= -8\gamma (n-2)\,a_{n-1}  +2\sum_{r=2}^{n-2}\, (r-1)(n-r-1)
  \,a_r\, a_{n-r}  \,. 
\end{equation}
Hence all the coefficients $a_n$ will be determined by these
equations. It also follows that these coefficients are real. The
reader may use this equation to find $a_4= -\tfrac{640}{3}\,\gamma^4$,
which can also be directly verified from
\eqref{eq:topological-string}. We will need this result shortly.

It is now straightforward to rewrite \eqref{eq:hol-anomaly-eq} as 
\begin{equation}
  \label{eq:diff-eq-h}
  \Big(\tilde\lambda\,\frac{\partial h_0}{\partial \tilde\lambda}\Big)^2 +
    \Big(\frac1{2\,\tilde\lambda} - 4\gamma\Big)
    \,\tilde\lambda\,\frac{\partial h_0}{\partial \tilde\lambda} +
    4\,\gamma^2 =0\,. 
\end{equation}
Since this equation is quadratic we may distinguish two different
solutions. One of them reproduces the weak coupling results,
\begin{equation}
  \label{eq:diff-eq-h-weak}
   \frac{\mathrm{d} h_0(\tilde\lambda)}{\mathrm{d}\tilde\lambda}=
   \frac{2\gamma}{\tilde\lambda} -\frac1{4\,\tilde\lambda^2} \bigg[1 -
   \sqrt{ 1-16\,\gamma\,\tilde\lambda}\;\bigg]\,. 
\end{equation}
This ordinary differential equation has a solution
\begin{equation} 
  \label{eq:sol-h}
  h_0(\tilde{\lambda}) = 2 \gamma \, \ln 4\gamma\tilde{\lambda} 
  + \frac{1}{4 \, \tilde{\lambda}}- \frac{\sqrt{1 
      - 16 \gamma \, \tilde{\lambda}}}{4 \, \tilde{\lambda}}
  -2\gamma + 4 \gamma \, 
  \operatorname{arctanh} \sqrt{1 - 16 \gamma \, \tilde{\lambda}}\;,
\end{equation}
whose expansion in powers of $\tilde\lambda$ indeed reproduces the first
three terms noted before,
\begin{equation}
 \label{eq:sol-h-approx}
 h_0(\tilde \lambda) = -8\,\gamma^2\tilde\lambda -
 32\,\gamma^3\,\tilde\lambda^2 -\frac{640}{3}\,
 \gamma^4\,\tilde\lambda^3 - 1792 \,  \gamma^5\,\tilde\lambda^4 +
 \mathcal{O}(\tilde\lambda^5)\,.
\end{equation}
Observe that this result for the STU-model is qualitatively different
from the result obtained in \cite{Alim:2015qma} for general Calabi-Yau
compactifications.
The function $h_0$ constitutes only part of the function $h$ and it is
not invariant under the STU-dualities. However, it is straightforward
to extend it to a duality invariant function by replacing the
definition of $\tilde\lambda$ according to
\begin{equation}
  \label{eq:new-lambda-tilde}
  \tilde \lambda\longrightarrow \tilde\lambda= - \,\frac{\lambda}{8
    \, \gamma^3}  \, D_S \, \omega(S) \; D_T \, 
    \omega(T) \; D_U \, \omega(U) \;,
\end{equation}
which reduces to the old definition \eqref{eq:eff-lambda} when
$\omega=0$.  Upon this replacement the function $h_0(\tilde\lambda)$
is STU-duality  invariant and the first few terms in its expansion are equal to
\begin{align}
  \label{eq:h0}
  h_0(\tilde \lambda) =\;& \sum_{n=2} a_n \,\tilde\lambda^{n-1}
  \nonumber \\
  =\;&\frac{\lambda}{\gamma} \big[D_S \, \omega(S) \; D_T \, 
    \omega(T) \; D_U \, \omega(U)\big]    -
       \frac{\lambda^2}{2\,\gamma^3}  \big[D_S \, \omega(S) \; D_T \,  
    \omega(T) \; D_U \, \omega(U)\big] ^2\nonumber\\
  &
  +\frac{5\,\lambda^3}{12\,\gamma^5} \big[D_S \, \omega(S) \; D_T \, 
    \omega(T) \; D_U \, \omega(U)\big]^3 \nonumber\\
  & - \frac{7\,\lambda^4}{16\,\gamma^7} 
    \big[D_S \, \omega(S) \; D_T \, 
    \omega(T) \; D_U \, \omega(U)\big]^4+\mathcal{O} (\lambda^5)\,, 
\end{align}
which is now manifestly duality invariant. Hence one can decompose
$h(\omega;\lambda)$ as follows,
\begin{equation}
  \label{eq:h-to-h0-h1}
    h(\omega;\lambda) = 
  \omega(S) + \omega(T) + \omega(U)  + h_0(\tilde \lambda) 
  + h_1 (\tilde\lambda) \;,
 \end{equation} 
 where the function $h_1(\tilde \lambda)$ should vanish in the limit
 $\omega=0$, because in that limit $h_0$ will already capture all the
 terms in $h$. Therefore $h_1$ can be written in a form that is at
 least linear in the covariant holomorphic functions $I^{(n)}$ defined
 in \eqref{eq:def-I}, or products thereof, which vanish for
 $\omega=0$, times first order covariant derivatives of the functions
 $\omega$. Explicit calculations leads to the first few terms in
 $h_1$,
\begin{align}
  \label{eq:h1-fct}
  h_1(\lambda) =& \, \frac{\lambda^2 }{2 \, \gamma^2} \Big[ I^{(2)}(S) \,
  (D_T \, \omega)^2 \, (D_U \, \omega)^2 + (D_S \, \omega)^2 \, I^{(2)}(T)
  \, (D_U \, \omega)^2 + (D_S \, \omega)^2 \, (D_T \, \omega)^2 \,
  I^{(2)}(U) \Big] \nonumber\\[1mm] 
  & + \frac{\lambda^3}{\gamma^3} \Big[
  (D_S \, \omega)^3 \, I^{(2)}(T) \, (D_T \, \omega) \, I^{(2)}(U) \, (D_U \,
  \omega) + I^{(2)}(S) \, (D_S \, \omega) \, \, I^{(2)}(T) \, (D_T \, \omega) \,
  (D_U \, \omega)^3  \nonumber\\ 
   & \qquad \quad + I^{(2)}(S) \, (D_S \,
  \omega) \, (D_T \, \omega)^3 \, I^{(2)}(U) \, (D_U \, \omega) \Big]
  \nonumber\\[1mm] 
&\ - \frac{\lambda^3}{ \, \gamma^4} \Big[ I^{(2)}(S) \, (D_S \, \omega ) \,
 (D_T \, \omega)^3 \, (D_U \, \omega)^3 +
(D_S \, \omega)^3 \, I^{(2)}(T) \,  (D_T \omega) \,
(D_U \, \omega)^3   \nonumber\\
& \qquad \quad 
+ (D_S \, \omega)^3 \,  (D_T \, \omega)^3 \, I^{(2)}(U) \, (D_U\, \omega)
\Big] \nonumber\\[1mm]
& + \frac{\lambda^3}{6 \, \gamma^3} \Big[ I{}^{(3)}(S) \,
  (D_T \, \omega)^3 \, (D_U \, \omega)^3 + (D_S \, \omega)^3 \, I{}^{(3)}(T)
  \, (D_U \, \omega)^3\nonumber\\
&\qquad \quad + (D_S \, \omega)^3 \, (D_T \, \omega)^3 \,
  I^{(3)}(U) \Big] + \mathcal{O}(\lambda^4) \,.
\end{align}
It is advantageous to express this result again in the modified
coupling constants $\tilde\lambda$ defined in
\eqref{eq:new-lambda-tilde}. Because of the invariance under dualities,
the expression takes a simpler form,
\begin{align}
  \label{eq:h1-fct}
  h_1(\tilde\lambda) =& \, 32\,\gamma^4 \big(\tilde\lambda^2+ 16\,
  \gamma\tilde\lambda^3\big) \Big[ \tilde {I}{}^{(2)}(S) \,
   + \tilde{I}{}^{(2)}(T)  + \tilde{I}{}^{(2)}(U) \Big] \nonumber\\[1mm] 
  & - 512\, \gamma^6 \tilde\lambda^3 \Big[
   \tilde{I}{}^{(2)}(T)  \,\tilde{I}{}^{(2)}(U)  +
 \tilde{I}{}^{(2)}(S) \,\tilde{I}{}^{(2)} (T) 
 + \tilde{I}{}^{(2)}(S) \,  \tilde{I}{}^{(2)}(U) \Big]
   \nonumber\\[1mm]
& - \frac{256}{3} \,\gamma^6\tilde\lambda^3 \Big[ \tilde{I}{}^{(3)}(S) \,
   + \tilde{I}{}^{(3)}(T) + \tilde{I}{}^{(3)}(U) \Big] + \mathcal{O}(\tilde\lambda^4) \,.
\end{align}
Here we made  use of the duality invariant quantities
\begin{equation}
  \label{eq:def-tilde-I}
  \tilde{I}^{(n)}(S) =  \frac{I^{(n)}(S)}{(D_S \,\omega)^n}\;, 
\end{equation}
which are no longer holomorphic because of the presence of the
non-holomorphic covariant derivative. Hence the function $h_1$ can be
written in terms of functions of $\tilde\lambda$ times polynomials of
the $\tilde{I}^{(n)}$ that are invariant under triality.

The function \eqref{eq:h-to-h0-h1} must still satisfy the anomaly
equation \eqref{eq:hol-anomaly-eq}, and this implies a non-linear
differential equation for $h_1$. To evaluate this equation we note the relations
\begin{align}
  \label{eq:der-lambda-tilde}
  \frac{\partial {\tilde \lambda}}{\partial {\bar S} }  =&\; \frac{2\,
    \gamma}{(S + \bar S)^2 \, D_S \, \omega}   \; {\tilde \lambda} \;,
  \nonumber\\[1mm]
  \frac{\partial \tilde{I}{}^{(n)} (S)}{\partial\bar S} =&\; -
  \frac{2\,n \,
    \gamma}{(S + \bar S)^2 \, D_S \, \omega}   \; \tilde{I}{}^{(n)}(S)  
  \nonumber\\[2mm]
  D_T \,  {\tilde \lambda} =&\;D_T\,\omega  \, \Big( \tilde{I}^{(2)}(T)
  -\frac{1}{2\gamma}\Big) \, {\tilde \lambda}
  \;,\nonumber\\[1mm]
  D_T \,\tilde{I}^{(n)}(T) =&\;  D_T\,\omega\, \Big( \tilde{I}^{(n+1)}(T)
  -\frac{n}{2\gamma} \tilde{I}^{(n)}(T) -n\,
  \tilde{I}^{(n)}(T)\,\tilde{I}^{(2)}(T) \Big) \,,
\end{align}
where we made use of \eqref{eq:I-n+1-nonholo-der} and
\eqref{eq:D2-omega}. 
With the help of these results one derives the equation for
$h_1(\tilde\lambda)$,
\begin{align}
  \label{eq:anom-eq-h1}
  & \tilde\lambda \frac{\partial h_1}{\partial\tilde\lambda} - \sum_{n=2}
  \,n\,\tilde{I}^{(n)}(S) \,\frac{\partial
    h_1}{\partial\tilde{I}^{(n)}(S) } =\\[1mm]
  &- 16\gamma^3\tilde\lambda\, \Big[ \tilde{I}^{(2)}(T) -\gamma\,
  \tilde{I}^{(2)}(T) \,\tilde{I}^{(2)}(U) + (T\leftrightarrow U)
  \Big]\nonumber\\[1mm]
  & -2\, \gamma\, \tilde\lambda \frac{\partial
    h_0}{\partial\tilde\lambda}\,\Big[ \big(1
  -4\,\gamma\tilde\lambda\big) \,\tilde{I}^{(2)}(T) - \gamma\big(1 -
    8\,\gamma\tilde\lambda \big)\, \tilde{I}^{(2)}(T) \,\tilde{I}^{(2)}(U)
  + (T\leftrightarrow U)\Big] \nonumber\\[1mm]
  & +4\, \gamma\,\tilde\lambda \, \Big[\Big( 1 -\frac1{2\gamma}
  \tilde\lambda \frac{\partial h_0}{\partial\tilde\lambda}\, \big( 1-
  2\gamma\,\tilde{I}^{(2)}(T) \big)\Big) \Big(1 -
  2\gamma\,\tilde{I}^{(2)}(U)\Big) + (T\leftrightarrow U) \Big]\,
  \tilde\lambda\frac{\partial
    h_1}{\partial\tilde\lambda} \nonumber\\[1mm]
  & -8\, \gamma^2\,\tilde\lambda \,\Big[\Big( 1- \frac1{2\gamma}
  \tilde\lambda \frac{\partial
    h_0}{\partial\tilde\lambda}\,\big(1-2\gamma\,\tilde{I}^{(2)}(T)\big)\Big)
  \nonumber\\ 
  &\qquad\qquad\quad  \times \sum_{n=2} \Big( \tilde{I}^{(n+1)}(U)
  -\frac{n}{2\gamma} \tilde{I}^{(n)}(U) -n\,
  \tilde{I}^{(n)}(U)\,\tilde{I}^{(2)}(U) \Big) 
   \,\frac{\partial
    h_1}{\partial\tilde{I}^{(n)}(U) }  + (T\leftrightarrow U) \Big]
  \nonumber\\
    &-8\gamma^2 \tilde\lambda\,  \nonumber\\
    &
    \times \Big[ \big( 1-  2\gamma\,\tilde{I}^{(2)}(T)\big)
    \frac1{2\gamma}
    \tilde\lambda \frac{\partial
    h_1}{\partial\tilde\lambda}- \sum_{m=2} \Big( \tilde{I}^{(m+1)}(T)
  -\frac{m}{2\gamma} \tilde{I}^{(m)}(T) -m\,
  \tilde{I}^{(m)}(T)\,\tilde{I}^{(2)}(T) \Big)   \,\frac{\partial
    h_1}{\partial\tilde{I}^{(m)}(T)}\Big]  \nonumber\\
&
\times \Big[ \big( 1-  2\gamma\,\tilde{I}^{(2)}(U)\big)\frac1{2\gamma}
  \tilde\lambda \frac{\partial
    h_1}{\partial\tilde\lambda}- \sum_{n=2}\Big( \tilde{I}^{(n+1)}(U)
  -\frac{n}{2\gamma} \tilde{I}^{(n)}(U) -n\,
  \tilde{I}^{(n)}(U)\,\tilde{I}^{(2)}(U) \Big)   \,\frac{\partial
    h_1}{\partial\tilde{I}^{(n)}(U)}\Big] \,, \nonumber
\end{align}
where we made use of \eqref{eq:diff-eq-h} without making use of the
weak coupling solution \eqref{eq:sol-h}. We will study the function
$h_1$ in the next section.

At the end of this section we analyze the more general solution of the
differential equation  \eqref{eq:diff-eq-h}. Because this 
 equation is quadratic, there are two branches,  
\begin{equation}
  \label{eq:diff-eq-h0}
   \frac{\mathrm{d} h_0(\tilde\lambda)}{\mathrm{d}\tilde\lambda}=
   \frac{2\gamma}{\tilde\lambda} -\frac1{4\,\tilde\lambda^2} \bigg[1 \mp
   \sqrt{ 1-16\,\gamma\,\tilde\lambda}\;\bigg]\,,
\end{equation}
which correspond to the two sheets of the Riemann surface described by
the following algebraic curve in $\mathbb{C}^2$,
\begin{equation}
  \label{eq:curve} 
  v^2 =  1-16\,\gamma\,\tilde\lambda \;,
\end{equation}
where $(\tilde\lambda, v) \in \mathbb{C}^2$. The first sheet is the
one that contains $(\tilde \lambda, v) = (0,1)$, while the second
sheet is the one that contains $(\tilde\lambda ,v) = (0,-1)$. The
weak coupling result derived so far is then recovered by working in
the first sheet in the vicinity of $(\tilde \lambda, v) = (0,1)$; the
associated ordinary differential equation \eqref{eq:diff-eq-h-weak} leads
to the solution \eqref{eq:sol-h}. Instead of working on one or on the other
sheet, we may work with a single variable $u$ on the Riemann surface
$\mathbb{C}$, which we identify with $v$ on the first sheet, and with
$-v$ on the second sheet. In terms of $u$ the function $h_0$ takes the
following form, 
\begin{equation}
  \label{eq:h0-u}
     h_0(u) = 2\gamma \Big(2\, \ln \frac{u+1}{2}  - \frac{u-1}{u+1} \Big) \;,
\end{equation}
and covers both sheets. Note that the solution $h_0(u)$ contains a
logarithmic branch cut starting at $u = -1$, and it vanishes at the
zero-coupling point $u=1$. The equation \eqref{eq:anom-eq-h1} for
$h_1$ depends on $h_0$ only through
\begin{equation}
  \label{eq:partial-lambda-conv}
  \tilde\lambda \,\frac{\partial h_0}{\partial\tilde\lambda} =
    2\gamma\,\frac{u-1}{u+1} \;,\qquad 
    \tilde\lambda \,\frac{\partial
      h_1}{\partial\tilde\lambda}= \frac{u^2-1}{2u} \,\frac{\partial
      h_1}{\partial u} \,, 
\end{equation}
where the relation between $\tilde\lambda$ and the duality invariant
effective coupling constant $u$ is given by
\begin{equation}
  \label{eq:u-lambdatilde}
  \frac{u^2-1}{16\gamma}  =- \tilde\lambda= 
  \frac{\lambda}{8\,\gamma^3} \,D_S\omega(S)\,D_T\omega(T)\,
  D_U\omega(U) \,.
\end{equation}

We now express equation \eqref{eq:anom-eq-h1} in terms of $u$, and we
will regard $h_1$ as a function of $u$,
\begin{align}
  \label{eq:anom-eq-h1-u}
  & \frac{u^2-1}{2u} \frac{\partial h_1}{\partial u} - \sum_{n=2}
  \,n\,\tilde{I}^{(n)}(S) \,\frac{\partial
    h_1}{\partial\tilde{I}^{(n)}(S) } =\\[1mm]
  &+ \gamma^2(u^2-1)\, \Big[ \tilde{I}^{(2)}(T) -\gamma\,
  \tilde{I}^{(2)}(T) \,\tilde{I}^{(2)}(U) + (T\leftrightarrow U)
  \Big]\nonumber\\[1mm]
  & - \gamma^2\,\frac{u-1}{u+1}
    \,\Big[ \big(3+u^2\big) \,\tilde{I}^{(2)}(T) - 2 \gamma 
    \big(1+u^2\big)\,
    \tilde{I}^{(2)}(T) \,\tilde{I}^{(2)}(U) 
  + (T\leftrightarrow U)\Big] \nonumber\\[1mm]
  & - \frac14 (u^2 -1) \, \Big[\Big( 1 -\frac{u-1}{u+1}\, \big( 1-
  2\gamma\,\tilde{I}^{(2)}(T) \big)\Big) \Big(1 -
  2\gamma\,\tilde{I}^{(2)}(U)\Big) + (T\leftrightarrow U) \Big]\,
  \frac{u^2-1}{2u} \frac{\partial h_1}{\partial u}\nonumber\\[1mm]
  & +\frac12 \gamma (u^2-1)  \,\Big[\Big( 1- \frac{u-1}{u+1}
  \,\big(1-2\gamma\,\tilde{I}^{(2)}(T)\big)\Big)
  \nonumber\\ 
  &\qquad\qquad\quad  \times \sum_{n=2} \Big( \tilde{I}^{(n+1)}(U)
  -\frac{n}{2\gamma} \tilde{I}^{(n)}(U) -n\,
  \tilde{I}^{(n)}(U)\,\tilde{I}^{(2)}(U) \Big) 
   \,\frac{\partial
    h_1}{\partial\tilde{I}^{(n)}(U) }  + (T\leftrightarrow U) \Big]
  \nonumber\\
    &+\frac12 \gamma (u^2-1)   \nonumber\\
    &
    \times \Big[ \big( 1-  2\gamma\,\tilde{I}^{(2)}(T)\big)
    \frac{u^2-1}{4 \gamma\, u} \frac{\partial h_1}{\partial u}
      - \sum_{m=2} \Big( \tilde{I}^{(m+1)}(T)
  -\frac{m}{2\gamma} \tilde{I}^{(m)}(T) -m\,
  \tilde{I}^{(m)}(T)\,\tilde{I}^{(2)}(T) \Big)   \,\frac{\partial
    h_1}{\partial\tilde{I}^{(m)}(T)}\Big]  \nonumber\\
&
\times \Big[ \big( 1-  2\gamma\,\tilde{I}^{(2)}(U)\big)
   \frac{u^2-1}{4\gamma\,u} \frac{\partial h_1}{\partial u}
  - \sum_{n=2}\Big( \tilde{I}^{(n+1)}(U)
  -\frac{n}{2\gamma} \tilde{I}^{(n)}(U) -n\,
  \tilde{I}^{(n)}(U)\,\tilde{I}^{(2)}(U) \Big)   \,\frac{\partial
    h_1}{\partial\tilde{I}^{(n)}(U)}\Big] \,.\nonumber
\end{align} 
In the next section we will discuss various partial solutions of this
equation. Note that we will be encountering three special values for
$u$, namely $u=1, 0, -1$.  The value $u=1$ corresponds to the
perturbative point. The significance of the other two points is at
present not entirely clear, as for those values the above equations
exhibit singularities.

\section{Evaluating contributions contained in  $h_1$}
\label{sec:addit-contr-h1}
\setcounter{equation}{0}
In this section we will start the explicit evaluation of a variety of
terms that are contained in the function $h_1$ by imposing the
holomorphic anomaly equation. This means that we will be studying
possible solutions of the differential equation
\eqref{eq:anom-eq-h1-u}. Since the function $h_1$ is decomposed in
terms of products of the quantities ${\tilde I}^{(n)}$ with
$u$-dependent coefficients, we can concentrate on specific products
and study the consequences of \eqref{eq:anom-eq-h1-u}.  As it turns
out, contributions that depend on at most two of the moduli are
determined by algebraic equations.  For the terms that depend on all
three moduli, the situation is more complicated since, for a subclass
of these terms, one will have to solve differential equations that
will necessarily introduce integration constants.

In the last part of this section we will then use the various
results of the topological string partition function and
investigate its implications for the effective Wilsonian
action. Although part of the input of the latter was taken into
account when constructing the former, it turns out that the dual
approach that we follow here enables not only to demonstrate the
mutual consistency of the corresponding results, but it also enables
to resolve the ambiguities that were encountered at this stage. For
instance, we will be able to fix an integration constant associated
with the differential equations that we will be trying to solve.

Let us first start by considering the terms in $h_1(u)$ that only depend
on $S$. In that case we derive the following equation from
\eqref{eq:anom-eq-h1-u},
\begin{align}
  \label{eq:only-I-S}
  \frac{u^2-1}{2} \,\frac{\partial h_1}{\partial u} -\sum_{n=2}
  \,n\,\tilde{I}^{(n)}(S) \,\frac{\partial
    h_1}{\partial\tilde{I}^{(n)}(S) }  -\frac1{32\,\gamma}
  \,\frac{(u^2-1)^3}{u^2} \,\Big(\frac{\partial h_1}{\partial
    u}\Big)^2=0 \,. 
\end{align}
However, there is a second equation that follows by interchanging $S$
and $T$ in \eqref{eq:anom-eq-h1-u}; here we also suppress all $T$-
and $U$-dependent terms, which leads to a different equation (because
the equation \eqref{eq:anom-eq-h1-u} is not symmetric under the
interchange of $S$ and $T$),
\begin{align}
  \label{eq:triality-u}
  &  2\,\gamma^2 \,\frac{(u-1)^2}{u+1} \,  \tilde{I}^{(2)}(S)
\nonumber\\ 
  &
  -\frac{u^2-1}{8u} \Big[ 3+u^2       - (u-3) (u-1) \, \big(1 -
  2\gamma\,\tilde{I}^{(2)}(S)\big) \Big] \,\frac{\partial h_1}{\partial u} \nonumber\\ 
  & + \Big( \gamma(u-1) - \frac{(u^2-1)^2}{8u} \,
  \frac{\partial h_1}{\partial u}\Big)   \sum_{n=2} \Big[ \tilde{I}^{(n+1)}(S)
  -\frac{n}{2\gamma} \tilde{I}^{(n)}(S) -n\,
  \tilde{I}^{(n)}(S)\,\tilde{I}^{(2)}(S) \Big] 
   \,\frac{\partial
    h_1}{\partial\tilde{I}^{(n)}(S) }  \nonumber\\ 
  & + \frac{1}{32\,\gamma} \, \big( 1-
  2\gamma\,\tilde{I}^{(2)}(S)\big)\,\frac{(u^2-1)^3}{u^2} 
  \,\Big(\frac{\partial h_1}{\partial u}\Big)^2=0 \,.
\end{align}
Combining this equation with \eqref{eq:only-I-S}, we derive an
equation that depends at most linearly on $\partial h_1/\partial u$,
\begin{align}
  \label{eq:triality-u-1}
  & \gamma\,\tilde{I}^{(2)}(S)\Big[ 2\,\gamma \,\frac{(u-1)^2}{u+1}   -
  \frac{(u^2-1)(u^2+3)}{4u}  \,\frac{\partial h_1}{\partial u} \Big]
  \nonumber\\
& + \Big( \gamma(u-1) - \frac{(u^2-1)^2}{8u} \,
  \frac{\partial h_1}{\partial u}\Big)   \sum_{n=2}  \tilde{I}^{(n+1)}(S)
  \,\frac{\partial
    h_1}{\partial\tilde{I}^{(n)}(S) }  \nonumber\\
& - \frac1{2} \big( 1+
  2\gamma\,\tilde{I}^{(2)}(S)\big)\,
  \Big( u-1  - \frac{(u^2-1)^2}{8\gamma \,u} \,
  \frac{\partial h_1}{\partial u}\Big)   \sum_{n=2} 
  n \,\tilde{I}^{(n)}(S)     \,\frac{\partial
    h_1}{\partial\tilde{I}^{(n)}(S) }  \nonumber\\ 
  &- \big( 1-
  2\gamma\,\tilde{I}^{(2)}(S)\big)\,  \sum_{n=2} n\, \tilde{I}^{(n)}(S) \,
  \frac{\partial h_1}{\partial\tilde{I}^{(n)}(S) }  =0\,. 
\end{align}
From the above equation one can then straightforwardly derive the
contributions to $h_1$ that are linearly proportional to
$\tilde{I}^{(n)}(S)$. The resulting expression, which also satisfies
\eqref{eq:only-I-S}, takes the form,
\begin{equation}
  \label{eq:h1-linear}
  h_1(u) \Big\vert_\mathrm{linear} = \sum_{m =2}  \, c_m(u)
  \,\big[ \tilde{I}^{(m)}(S)   +\tilde{I}^{(m)}(T) +\tilde{I}^{(m)}(U)\big]  \,,
\end{equation}
with 
\begin{equation}
  \label{eq:c-m}
  c_m(u) = \frac{(2\gamma)^m}{m!}  \,\Big(\frac{u-1}{u+1}\Big)^m\,.
\end{equation}
Note that we included the $T$- and $U$-dependent terms in
\eqref{eq:h1-linear} to make the result manifestly invariant under
triality.

Encouraged by this result, we proceed to determine the coefficient
functions of the terms in $h_1$ equal to
$\tilde{I}^{(m)}(S)\,\tilde{I}^{(n)}(S)$ as well as
$\tilde{I}^{(m)}(S)\,\tilde{I}^{(n)}(T)$,
\begin{align}
  \label{eq:h1-quadratic} 
  h_1(u) \Big\vert_\mathrm{quadratic} = \sum_{m,n =2}  \,&  d_{m,n}(u)
  \, \big[ \tilde{I}^{(m)}(S)\,\tilde{I}^{(n)}(S) +
  \tilde{I}^{(m)}(T)\,\tilde{I}^{(n)}(T)
  +\tilde{I}^{(m)}(U)\,\tilde{I}^{(n)}(U)\big] \\
   & +  \,e_{m,n} (u) \,\big[
   \tilde{I}^{(m)}(S)\,\tilde{I}^{(n)}(T) +
   \tilde{I}^{(m)}(T)\,\tilde{I}^{(n)}(U) +
   \tilde{I}^{(m)}(U)\,\tilde{I}^{(n)}(S) \big]   \,, \nonumber
\end{align}
where we included the terms related by triality. Obviously
$d_{m,n}(u)$ and $e_{m,n}(u)$ are symmetric in $(m,n)$.  The
contributions \eqref{eq:h1-quadratic} will be determined from
\eqref{eq:anom-eq-h1-u}, which leads to three different equations. The
first one is equal to
\begin{align}
  \label{eq:only-S-T-1}
  & \; \frac{u^2-1}{2} \,\frac{\partial h_1}{\partial u} 
  -\frac1{32\,\gamma} 
  \,\frac{(u^2-1)^3}{u^2} 
  \,\Big(\frac{\partial h_1}{\partial u} \Big)^2 
  \,\big(1-2\gamma \,\tilde{I}^{(2)} (T)\big) \nonumber\\
  &  -  \sum_{n=2} n\, \tilde{I}^{(n)}(S) \,
  \frac{\partial h_1}{\partial\tilde{I}^{(n)}(S)} 
  +  \gamma\,(u-1)(u-3) \, \frac{u^2-1} {4u} \, \frac{\partial
    h_1}{\partial u} \,\tilde{I}^{(2)}(T)
  \nonumber\\
 & + \frac1{2\gamma} 
 \Big[ \gamma(u-1)   
   -\frac{(u^2-1)^2}{8\,u} \,\frac{\partial h_1}{\partial u}  \Big]
   \, \big( 1+ 2\gamma\,\tilde{I}^{(2)}(T)\big) \,
  \sum_{n=2}  n\,\tilde{I}^{(n)}(T)
  \,\frac{\partial
    h_1}{\partial\tilde{I}^{(n)}(T) }  \nonumber\\
  & - \bigg[ \gamma(u-1)  
   - \frac{(u^2-1)^2}{8\,u} \,\frac{\partial h_1}{\partial
    u} \Big] \, 
  \sum_{n=2}  \tilde{I}^{(n+1)}(T)
  \,\frac{\partial
    h_1}{\partial\tilde{I}^{(n)}(T) }   =0 \,,
\end{align}
where we have retained all the terms depending on $S$ and $T$ with the
exception of a term linear in $\tilde{I}^{(2)}(T)$. A second equation
follows from exchanging $S\leftrightarrow T$ in \eqref{eq:anom-eq-h1-u},
suppressing all the $U$-dependent term. This will lead to equation
\eqref{eq:only-S-T-1} with $S$ and $T$ interchanged.  Finally the
third equation follows from interchanging $S$ and $U$ in
\eqref{eq:anom-eq-h1-u}, and subsequently suppressing all terms that
depend on $U$; this equation is symmetric in $S$ and $T$,
\begin{align}
  \label{eq:only-S-T-3}
  &  \frac{(u^2-1)}{2}\, \frac{\partial h_1}{\partial u}   -
  2\gamma^3 \frac{(u-1)^3}{u+1}  \,\tilde{I}^{(2)}(S)\,
  \tilde{I}^{(2)} (T) 
  \\[1mm]
  & + \gamma\,
  \frac{u^2-1}{4u} \Big[ (u-3)(u-1)\, 
  \big(\tilde{I}^{(2)}(S)+\tilde{I}^{(2)} (T) \big) - 4\gamma (u-1)^2
  \,\tilde{I}^{(2)}(S)\,\tilde{I}^{(2)} (T)  \Big] \,\frac{\partial
    h_1}{\partial u} 
  \nonumber\\[1mm]
  &-\gamma(u-1) \,\sum_{n = 2} \bigg[\Big[(1+\gamma(u-1)
  \,\tilde{I}^{(2)} (S)\big) \,\tilde{I}^{(n+1)}(T)
  \,\frac{\partial h_1}{\partial\tilde{I}^{(n)}(T)} +
  \big(S\leftrightarrow T\big) \Big]
  \nonumber\\[1mm]
  &\qquad\qquad - \frac1{2\gamma} 
  \Big[\big(1+\gamma(u-1)
  \,\tilde{I}^{(2)} (S)\big) \,\big(1+2\gamma\,\tilde{I}^{(2)}
  (T)\big)\,   n\, \tilde{I}^{(n)}(T) 
  \,\frac{\partial h_1}{\partial\tilde{I}^{(n)}(T)} 
  +\big(S\leftrightarrow T\big) \Big] \bigg]
  \nonumber\\[2mm]
  &   -\frac1{32\,\gamma} 
  \,\frac{(u^2-1)^3}{u^2} 
  \,\Big(\frac{\partial h_1}{\partial u} \Big)^2 
  \,\big(1-2\gamma \,\tilde{I}^{(2)} (S)\big)\,\big(1-2\gamma
  \,\tilde{I}^{(2)} (T)\big)  
  \nonumber\\[1mm]
  & +\frac{(u^2-1)^2} {8u} \,\frac{\partial h_1}{\partial u}
  \,\sum_{n = 2} \bigg[\Big[ 
  \big(1-2\gamma \,\tilde{I}^{(2)} (S)\big)\, \tilde{I}^{(n+1)}(T)
  \,\frac{\partial h_1}{\partial\tilde{I}^{(n)}(T) } +
  \big(S\leftrightarrow T\big)\Big] \nonumber\\[1mm]
  &\qquad\qquad -\frac{1}{2\gamma}\Big[  \big(1-2\gamma
    \,\tilde{I}^{(2)} 
  (S)) \,\big(1+2\gamma \,\tilde{I}^{(2)} (T))\, 
   n\,  \tilde{I}^{(n)}(T) \,\frac{\partial
     h_1}{\partial\tilde{I}^{(n)}(T) }   +
  \big(S\leftrightarrow T\big) \Big]  \bigg] 
  \nonumber\\[2mm]
  &-\tfrac1{2} \gamma (u^2-1) \sum_{m,n =  2} \bigg[ \tilde{I}^{(m+1)}(S)
  \,\frac{\partial h_1}{\partial\tilde{I}^{(m)}(S) } \,\tilde{I}^{(n+1)}(T)
  \,\frac{\partial h_1}{\partial\tilde{I}^{(n)}(T) } 
  \nonumber\\[2mm]
  & \qquad\qquad -\frac{1}{2\gamma} \Big[\big(1+2\gamma
    \,\tilde{I}^{(2)} 
  (T)) \,\tilde{I}^{(m+1)}(S) \,\frac{\partial
    h_1}{\partial\tilde{I}^{(m)}(S) } \,n\, \tilde{I}^{(n)}(T)
  \,\frac{\partial h_1}{\partial\tilde{I}^{(n)}(T) } +
  \big(S\leftrightarrow T\big)\Big] \nonumber\\[1mm]
   &\qquad\qquad + \frac1{4\gamma^2}\big(1+2\gamma \,\tilde{I}^{(2)}(S)\big) \,
   \big(1+2\gamma \,\tilde{I}^{(2)} (T)\big)\, m\, \tilde{I}^{(m)}(S)
  \,\frac{\partial h_1}{\partial\tilde{I}^{(m)}(S) }\,n\,\tilde{I}^{(n)}(T)
  \,\frac{\partial h_1}{\partial\tilde{I}^{(n)}(T) } \bigg]
  = 0 \,. \nonumber
\end{align}
Here we have dropped one term linear in
$\big(\tilde{I}^{(2)}(S)+ \tilde{I}^{(2)}(T)\big)$, since we are interested in determining the quadratic terms in \eqref{eq:h1-quadratic}.

When considering only terms of second order in the $\tilde{I}^{(n)}$,
the equation \eqref{eq:only-S-T-1} simplifies to
\begin{align}
  \label{eq:only-S-T-1-cont}
  & \; \frac{u^2-1}{2} \,\frac{\partial h_1}{\partial u} 
  -\frac1{32\,\gamma} 
  \,\frac{(u^2-1)^3}{u^2} 
  \,\Big(\frac{\partial h_1}{\partial u} \Big)^2 \nonumber\\
  &
    -  \sum_{n=2} \bigg[ n\,\Big[ \tilde{I}^{(n)}(S) \,
    \frac{\partial h_1}{\partial\tilde{I}^{(n)}(S)} - \tfrac12 (u-1)
    \tilde{I}^{(n)}(T) \, 
    \frac{\partial h_1}{\partial\tilde{I}^{(n)}(T)}  \Big]
    +\gamma(u-1)  \tilde{I}^{(n+1)}(T)
    \,\frac{\partial h_1}{\partial\tilde{I}^{(n)}(T) }  \bigg]
\nonumber\\
  & =   -  \gamma\,(u-1)(u-3) \, \frac{u^2-1} {4u} \, \frac{\partial
    h_1}{\partial u} \,\tilde{I}^{(2)}(T)
  \nonumber\\
 &\quad   
   + \frac{(u^2-1)^2}{16\gamma\,u} \,\frac{\partial h_1}{\partial u}   
  \,\sum_{n=2} \, \big[ n\,\tilde{I}^{(n)}(T)  -2\gamma\,
   \tilde{I}^{(n+1)}(T)\big] 
  \,\frac{\partial h_1}{\partial\tilde{I}^{(n)}(T) }  \nonumber\\
  &\quad
    -\gamma (u-1) \tilde{I}^{(2)}(T)  
  \,\sum_{n=2}  n\,\tilde{I}^{(n)}(T)
  \,\frac{\partial
    h_1}{\partial\tilde{I}^{(n)}(T) }      \,.
\end{align}
The terms on the right-hand side of this equation yield all the terms
quadratic in $\tilde{I}^{(n)}$ and are determined by
\eqref{eq:h1-linear}.

Also equation \eqref{eq:only-S-T-3} can be simplified by suppressing
all terms that manifestly lead to third and higher orders in
$\tilde{I}^{(n)}$, 
\begin{align}
  \label{eq:only-S-T-3-cont}
  &  \frac{(u^2-1)}{2}\, \frac{\partial h_1}{\partial u}   
    -\frac1{32\,\gamma}   \,\frac{(u^2-1)^3}{u^2} 
    \,\Big(\frac{\partial h_1}{\partial u} \Big)^2 \nonumber\\
  &
    + \tfrac12  (u-1) \sum_{n=2} n \Big[ \tilde{I}^{(n)}(S) \,
    \frac{\partial h_1}{\partial\tilde{I}^{(n)}(S)} +
    \tilde{I}^{(n)}(T) \,
    \frac{\partial h_1}{\partial\tilde{I}^{(n)}(T)}\Big] \nonumber\\
  & -\gamma (u-1) \sum_{n=2} \,\Big[ \tilde{I}^{(n+1)}(S)
    \,\frac{\partial h_1}{\partial\tilde{I}^{(n)}(S) } +  \tilde{I}^{(n+1)}(T)
    \frac{\partial h_1}{\partial\tilde{I}^{(n)}(T)} 
    \Big]   \nonumber\\
  &=  2\gamma^3 \frac{(u-1)^3}{u+1}  \,\tilde{I}^{(2)}(S)\,
  \tilde{I}^{(2)} (T) \nonumber\\[1mm]
  &\quad - \gamma\,
  \frac{u^2-1}{4u} \Big[ (u-3)(u-1)\, 
  \big(\tilde{I}^{(2)}(S)+\tilde{I}^{(2)} (T) \big) \Big] \,\frac{\partial
    h_1}{\partial u} 
  \nonumber\\[1mm]
  &\quad +\gamma(u-1) \,\sum_{n = 2} \bigg[\Big[ \gamma(u-1)
  \,\tilde{I}^{(2)} (S) \, \tilde{I}^{(n+1)}(T)
  \,\frac{\partial h_1}{\partial\tilde{I}^{(n)}(T)} +
  \big(S\leftrightarrow T\big) \Big]
  \nonumber\\[1mm]
  &\qquad\qquad\qquad\qquad - \frac1{2} 
  \Big[\big((u-1)
  \,\tilde{I}^{(2)} (S) +2\tilde{I}^{(2)}
  (T)\big)\,   n\, \tilde{I}^{(n)}(T) 
  \,\frac{\partial h_1}{\partial\tilde{I}^{(n)}(T)} 
  +\big(S\leftrightarrow T\big) \Big] \bigg]
  \nonumber\\[2mm]
  &\quad 
    -\frac{(u^2-1)^2} {8u} \,\frac{\partial h_1}{\partial u}
  \,\sum_{n = 2} \bigg[  \tilde{I}^{(n+1)}(T)
  \,\frac{\partial h_1}{\partial\tilde{I}^{(n)}(T) } 
   -\frac{1}{2\gamma}  n\,  \tilde{I}^{(n)}(T) \,\frac{\partial
     h_1}{\partial\tilde{I}^{(n)}(T) }   +
  \big(S\leftrightarrow T\big)  \bigg] 
  \nonumber\\[1mm]
  &\quad   +\tfrac1{2} \gamma (u^2-1) \sum_{m,n =  2} \bigg[ \tilde{I}^{(m+1)}(S)
  \,\frac{\partial h_1}{\partial\tilde{I}^{(m)}(S) } \,\tilde{I}^{(n+1)}(T)
  \,\frac{\partial h_1}{\partial\tilde{I}^{(n)}(T) } 
  \nonumber\\[2mm]
  &\quad \qquad\qquad\qquad\qquad -\frac{1}{2\gamma} \Big[ 
   \,\tilde{I}^{(m+1)}(S) \,\frac{\partial
    h_1}{\partial\tilde{I}^{(m)}(S) } \,n\, \tilde{I}^{(n)}(T)
  \,\frac{\partial h_1}{\partial\tilde{I}^{(n)}(T) } +
  \big(S\leftrightarrow T\big)\Big] \nonumber\\[1mm]
   &\quad\qquad\qquad\qquad\qquad + \frac1{4\gamma^2}\, m\,
     \tilde{I}^{(m)}(S) 
  \,\frac{\partial h_1}{\partial\tilde{I}^{(m)}(S) }\,n\,\tilde{I}^{(n)}(T)
  \,\frac{\partial h_1}{\partial\tilde{I}^{(n)}(T) } \bigg] \;.
\end{align}
Just as in \eqref{eq:only-S-T-1-cont}, the terms on the right-hand side of
this equation yield all the terms quadratic in $\tilde{I}^{(n)}$ and are
determined by \eqref{eq:h1-linear}.

Note that the first lines of \eqref{eq:only-S-T-1-cont} and of
\eqref{eq:only-S-T-3-cont} are identical.  Hence, by taking the
difference between these two equations we obtain an equation that does
not contain derivatives of the coefficient functions $d_{m,n}(u)$ and
$e_{m,n}(u)$, and that relates them to the coefficient functions
$c_m(u)$ given in \eqref{eq:c-m}.  By comparing terms with the same
powers of $\tilde{I}^{(m)}(S) \tilde{I}^{(n)}(S)$ or
$\tilde{I}^{(m)}(S) \tilde{I}^{(n)}(T)$, we determine the explicit
form of the coefficient functions $d_{m,n}(u)$ and $e_{m,n}(u)$.  Let
us illustrate this by focussing on the first two functions
$d_{2,2}(u)$ and $e_{2,2} (u)$.  For the coefficient function
$d_{2,2}(u)$ we obtain the relation
\begin{eqnarray}
  \label{d22-1}
   d_{2,2}(u) =  
  \frac{(u^2 -1)^2}{16 \gamma \, u(u+1) } \, c_2(u) \,\dot{c}_2(u) 
  - \gamma \, \frac{(u -1)^2 (u-3)}{8 u}  \, \dot{c}_2(u) 
  -  \gamma  \, \frac{u-1}{u+1} \, c_2(u)  \;, 
\end{eqnarray}
where here and henceforth we will use the notation $\dot{c}(u)=
\mathrm{d}c(u)/\mathrm{d}u$. Substituting \eqref{eq:c-m},  we find
\begin{equation}
  \label{eq:d22}
  d_{2,2}(u) = - (2 \gamma)^3 \, \frac{(u+2)}{4u} \left( \frac{u-1}{u+1}
  \right)^4 \;. 
\end{equation}
For the coefficient function $e_{2,2} (u)$ we derive the relation
\begin{align}
  \label{eq:e22-1}
   e_{2,2}(u) =&\,
  \frac{(u^2 -1)^2}{8 \gamma \, u(u+1)} \, c_2(u) \,\dot{c}_2(u) 
   + \frac{u-1)}{2 \gamma}  \, c_2(u){}^2
    - \frac{\gamma \,  (u -1)^2(u-3)}{4 u}  \, \dot{c}_2(u) \nonumber\\
    &\, - 2 \gamma \,\frac{ (u-1)^2}{u+1} \, c_2(u)  
      + 2 \, \gamma^3 \, \frac{(u-1)^3}{u+1} \;,
\end{align}
which, upon using again \eqref{eq:c-m},  yields
\begin{equation}
  \label{eq:e22}
  e_{2,2}(u) = \frac{(2\,\gamma)^3}{u} \, \Big( \frac{u-1}{u+1} \Big)^3 \;.
\end{equation}
It can be verified that these expressions for $d_{2,2}(u) $ and $e_{2,2}(u) $
also solve \eqref{eq:only-S-T-1}.

Proceeding in this manner, we obtain the following expressions for the
coefficient functions (with $m,n\geq 2$),
\begin{align}
  \label{eq:de-(mn)}
  d_{m,n}(u) =&\; -\frac{(2\gamma)^{m+n-1}}{(m-1)!\, (n-1)!} \, 
  \bigg[ \frac{1}{m(m+1)}
  +\frac{1}{n(n+1)}  - \frac{(m+n-2)!}{(m+n)!} +\frac{1}{2 u} \,
  \bigg]\,   \Big(\frac{u-1}{u+1}\Big)^{m+n} \,,
   \nonumber\\[1mm]
    e_{m,n} (u) =&\; \frac{ (2 \gamma)^{m + n -1}}{(m-1)! \,
    (n-1)!  } \,  \frac1{u} \,\Big( \frac{u-1}{u+1} \Big)^{m + n -1}\,. 
\end{align}
Note that these functions exhibit singularities at $u=0$ and $u=-1$.

It is remarkable that the coefficient functions $d_{m,n}(u)$ and
$ e_{m,n}(u) $ are determined by algebraic means, while we started
from a differential equation. This is related to the fact that the
equation \eqref{eq:anom-eq-h1-u} that we are trying to solve, is not
symmetric under the interchange of the moduli. To be more precise,
when considering the coefficient function belonging to a given product
of powers of the symmetric combinations
$\big[\tilde{I}^{(n)}(S) \,\tilde{I}^{(n)} (T)
\,\tilde{I}^{(n)}(U)\big]$, the derivative terms cannot be removed and
one has to solve a differential equation, which will lead to 
integration constants. However, when considering structures that are
less symmetric, one can eliminate the $u$-derivatives on $h_1$ by
anti-symmetrizing in two of the moduli and find suitable algebraic
equations.

We will demonstrate that differential equations cannot altogether be
avoided by discussing two examples, whose results are needed at the
end of this section when we will try to match the results of this
section to the explicit expression obtained in section
\ref{sec:higher-orders} for the holomorphic function that encodes the
Wilsonian action. The first example concerns a coefficient function
denoted by $p(u)$, that can be determined algebraically.  It appears
in $h_1$ as
\begin{align}
  \label{eq:p(u)}
    h_1(u) = p(u)\, \big[ (\tilde{I}^{(2)} (S) )^2\, \big( \tilde{I}^{(2)}(T)
    +\tilde{I}^{(2)}(U)\big)  &+  (\tilde{I}^{(2)} (T) )^2 \big(\tilde{I}^{(2)}(U) 
       +\tilde{I}^{(2)}(S)\big) 
    \nonumber\\
   &+ (\tilde{I}^{(2)} (U) )^2 \big(\tilde{I}^{(2)}(S)
     +\tilde{I}^{(2)}(T)\big)\big]+ \cdots \,.
\end{align}
Proceeding as before, we consider \eqref{eq:anom-eq-h1-u} and subtract
from it the same equation with $S$ and $T$ interchanged. The difference
between these two equations gives rise to an algebraic equation for
$p(u)$, 
\begin{align}
    \label{eq:algp}
  p(u) =&\; \frac{\gamma (u-1)^2(u-3)}{4\,u}  \big[ \dot{d}_{2,2}  
        - \dot{e}_{2,2} 
          \big] - 2\gamma \, \frac{(u-1)}{(u+1)} \, e_{2,2}  
          - \frac{u-1}{4} \Big( \frac{u^2-1}{2u} \, \dot{c}_2  \Big)^2 \nonumber\\
        &+  \frac{(u-1)^2 (u+1)}{8\gamma\,u}
          \big[ - c_2
          \big( \dot{d}_{2,2}  - \dot{e}_{2,2} \big)
          + 2 d_{2,2}  \, \dot{c}_2 + 2\, \gamma \, c_2 \, \dot{c}_2
          \big] \,.
\end{align}
It is easy to see that $p(u)$ is proportional to $(u-1)^4$. The
leading contributions  originate from the terms $\dot{e}_{2,2(u)}$ and
$e_{2,2}(u)$ in the first line. They lead to 
\begin{equation}
  \label{eq:pu4}
  p(u) = \tfrac12\gamma^4\, (u-1)^4 +\mathcal{O}\big((u-1)^5\big) \,.
\end{equation}
 
The second example, which does not lead to an algebraic equation,
concerns the following terms,
\begin{align}
  \label{eq:exp_h1_rel}
  h_1(u) =&\,  c_2(u) \, \big[ \tilde{I}^{(2)}(S) + \tilde{I}^{(2)}(T) + 
             \tilde{I}^{(2)}(U) \big] \nonumber\\
           &\,+ e_{2,2}(u) \,\big[  \tilde{I}^{(2)}(S)\,
            \tilde{I}^{(2)}(T) +  
             \tilde{I}^{(2)}(T) \,\tilde{I}^{(2)}(U)
             + \tilde{I}^{(2)}(U) \,\tilde{I}^{(2)}(S) \big] \nonumber\\
           &\, + v_{2,2,2}(u) \, \tilde{I}^{(2)}(S)
             \,\tilde{I}^{(2)}(T) \,
              \tilde{I}^{(2)}(U) + \cdots \,.
\end{align}
Inserting \eqref{eq:exp_h1_rel} into \eqref{eq:anom-eq-h1-u} leads to
the differential equation,
\begin{align}
\label{eq:diff-v}
\frac{(u^2 -1)}{2} \,& \dot{v}_{2,2,2} - 2 (2 -u) \,v_{2,2,2}  \nonumber\\
   =&\,\frac{(u^2-1)^2}{4u} \Big[ 2 \gamma \, \dot{e}_{2,2} - 
    4 \gamma \,\dot{e}_{2,2} \,  \frac{u-1}{u+1} + 4 \gamma^2\,
      \dot{c}_2  \,  
    \left( \frac{u-1}{u+1} \right) \Big]  
    - 2 \gamma \, e_{2,2} \, (u -1)^2 \nonumber\\[1mm]
      &\,  + \frac{(u^2 -1)}{2 \gamma} 
        \Big[ \frac{(u^2 -1)^2}{16 u^2} \left( 6\, \dot{e}_{2,2}\, \dot{c}_2 - 8
        \gamma\, (\dot{c}_2)^2\right)  \nonumber\\
      &\,  \qquad \qquad \qquad 
        + \frac{(u^2 -1)}{2 u} \left( \dot{e}_{2,2}\, c_2 + 2\, e_{2,2} \,\dot{c}_2 
        - 2 \gamma \,\dot{c}_2\, c_2\right) + 2\, e_{2,2}\, c_2  \Big] \,.
\end{align}
Evaluating the right-hand side of this expression by using the values
for $c_2(u)$ and $e_{2,2}(u)$ given in \eqref{eq:c-m} and \eqref{eq:d22},
one obtains
\begin{equation}
    - \frac{4 \gamma^4 }{u^4}  \left( \frac{u-1}{u+1} 
    \right)^4 (u^2 +1)
    \left( 3(u^2 -1) - 16\,u \right) \;.  
\end{equation}
The differential equation \eqref{eq:diff-v} for $v_{2,2,2}(u)$ is then
solved by 
\begin{equation}
  \label{eq:sol-v-alp}
  v_{2,2,2}(u) = 8\gamma^4 \Big(\frac{u-1}{u+1}\Big)^2\,F(u)\,,
\end{equation}
with 
\begin{equation}
    \label{eq:F-u}
    F(u) = \frac{1}{(u+1)^4}\, \Big[ \alpha_v - 
  \frac{ u^6 -1  - 8u (u^4 + 1) - 3 u^2 (u^2 -1)} {u^3} \,\Big] \;,
\end{equation}
where $\alpha_v$ denotes an integration constant. 

At this point we have determined quite a number of coefficient
functions in $h_1(u)$. To obtain the complete result for the
topological string partition function, they have to be combined with
the contributions from the function $h_0(\tilde\lambda)$, which was
obtained by resummation and is thus known to all orders in
$\tilde\lambda$. An interesting question is whether one can now
re-obtain the result for the holomorphic function that encodes the
Wilsonian action, and if so, whether this procedure will have
implications for some of the ambiguities that we encountered in
section \ref{sec:higher-orders}. We will now demonstrate that this is
indeed the case.

As we argued in the text above \eqref{eq:hole-limit-h}, one can in
principle match the results for the topological string partition
function with those for the Wilsonian action. The latter is expressed in terms of the
holomorphic functions $\omega^{(n)}$ that, for $n\leq5$, were
determined in section \ref{sec:higher-orders}. To do so, note that the
coefficient functions of $h_1(u)$ have to be expressed in terms
$\tilde\lambda$, which can be achieved by first expressing the duality
invariant coupling $u$ in terms of $\tilde\lambda$ by an expansion
about the perturbative point $u=1$,
\begin{equation}
  \label{eq:u-expansion}
  u= 1-8\gamma \,\tilde{\lambda} - 32\gamma^2\,\tilde\lambda^2 +
  \mathcal{O}(\tilde\lambda^3)\,.
\end{equation} 
After this conversion one writes $\tilde\lambda$ in terms of
$\lambda$, the parameter used in section
\ref{sec:STU-model-dualities}, making use of equation
\eqref{eq:new-lambda-tilde}, and substitutes the explicit expression
for the quantities $\tilde{I}^{(n)}$. Finally one suppresses all the
(non-holomorphic) connections in the covariant derivatives $D\omega$
with respect to the moduli, by replacing $D\omega$ with simple
derivatives $\partial\omega$. The result of this truncation can then
be compared to the functions $\omega^{(n)}$ with $n\leq5$. The terms
belonging to $\omega^{(n)}$ can then simply be identified because they
will be proportional to $\lambda^{(n-1)}$, as follows from the
expansion \eqref{eq:Ups-holo-expansion}. We will now briefly review
the results of this analysis.

We first note that according to the procedure sketched above, the
function $h_0$ will take the form of a power series in
$\lambda\,\partial_S\omega\, \partial_T\omega\,\partial_U\omega$,
where $\lambda$ denotes the original expansion parameter in section
\ref{sec:higher-orders}. In equation \eqref{eq:h0} we have recorded
the first four terms of this expansion, where we may now drop the
(non-holomorphic) connections in the covariant derivatives. Such terms
also appear in the functions $\omega^{(n)}$ that encode the Wilsonian
effective action, but it is easy to see that the terms in
$\omega^{(2)}$- $\omega^{(5)}$ do not agree with the first four terms
in \eqref{eq:h0}. Hence the contributions from $h_1(u)$ should
compensate for this difference.  Indeed, such terms are present
in $h_1(u)$, and they originate from those terms in the quantities
$\tilde{I}^{(n)}$ that are proportional to a constant. For convenience
we list the values of these constants for $n=2,3,4$, which follow
directly from \eqref{eq:def-I},
\begin{equation}
  \label{eq:2-constant}
  \tilde{I}^{(2)}_\text{constant}= \frac1{2\gamma} \,,\quad
  \tilde{I}^{(3)}_\text{constant} = \frac1{\gamma^2}\,, 
  \quad \tilde{I}^{(4)} _\text{constant}=
  \frac3{\gamma^3} \,.
\end{equation}

Let us now identify the contributions from $h_1(u)$ that will contribute
to terms that are at most of fourth order in $(u-1)$. They consist of
the terms proportional to $c_2(u)$, $c_3(u)$ and $c_4(u)$ in
\eqref{eq:h1-linear},  the terms proportional to $d_{2,2}(u)$,
$e_{2,2}(u)$ and $e_{2,3}(u)$ in \eqref{eq:h1-quadratic}, and finally
the terms proportional to the coefficient functions $p(u)$ and
$v_{2,2,2}(u)$ that appear in \eqref{eq:p(u)} and
\eqref{eq:exp_h1_rel}, respectively. 

In principle all these contributions may contribute to the
functions $\omega^{(2)}$- $\omega^{(5)}$ determined in sections
\ref{sec:STU-model-dualities} and \ref{sec:higher-orders}. Let us
first concentrate on the constant part of the functions
$\tilde{I}^{(n)}$ that will combine with the  terms of
$h_0(\tilde\lambda)$. In that case the relevant contributions from
$h_1(u)$ will take the form,
\begin{align}
  \label{eq:h1-constant}
  h_1(u)\big|_\text{constant} =\;& \frac{3}{2\gamma}\,c_2(u)  
         + \frac{3}{\gamma^2}\, c_3(u)
        + \frac{9}{\gamma^3} \,c_4(u) 
     + \frac{3}{4\gamma^2} \big[d_{2,2}(u) + e_{2,2}(u) \big]\nonumber\\
     &+ \frac{3}{\gamma^3}\, e_{2,3}(u)
      +\frac{1}{8\gamma^3} \big[ 6\,p(u) + v_{2,2,2}(u) \big] \cdots \,.
\end{align}
From the explicit expressions of the coefficient functions one can
easily determine that only the first term proportional to $c_2(u)$ and
the term proportional to $v_{2,2,2}(u)$ will contain terms of second
order in $(u-1)$, depending on the value of the integration constant
$\alpha_v$. Upon using \eqref{eq:u-expansion} it follows that the
corresponding contribution from $c_2(u)$ to $h_1(u)$ is equal to
$48\,\gamma^3 \tilde\lambda^2$. There is also a similar contribution
from $h_0(\tilde\lambda)$ which, according to \eqref{eq:sol-h-approx},
equals $-32\, \gamma^3 \tilde\lambda^2$, so that sum of these two
contributions yields $16\,\gamma^3\tilde\lambda^2$. Converting
$\tilde\lambda$ to $\lambda$ by using \eqref{eq:new-lambda-tilde},
and suppressing the connections in the covariant derivative, one
reproduces exactly the first term in the expression for $\omega^{(3)}$
given in \eqref{eq:higher-orders-3}. Therefore it follows that the
function $F(u)$ must vanish for $u=1$, which fixes the integration
constant $\alpha_v$ to
\begin{equation}
  \label{eq:value-alpha}
\alpha_v = - 16 \,.
\end{equation}
With this result $v_{2,2,2}(u)$ reads,
\begin{equation}
   \label{eq:v-222}     
   v_{2,2,2}(u) = - \frac{8 \gamma^4 }{u^3}  
   \left( \frac{u-1}{u+1} \right)^4 \left( u^2 -1 - 8u \right) \,.
\end{equation}
 
We can now determine all terms in \eqref{eq:h1-constant}, so that one
can evaluate all the contributions up to order ${\tilde \lambda}^5$,
leading to
\begin{align}
  \label{eq:h1-constant-final}
  h_1(u)\big|_\text{constant} =\;& 48 \gamma^3 \, {\tilde \lambda}^2
          + 128 \gamma^4 \, {\tilde \lambda}^3 
           + 2432 \gamma^5 \, {\tilde \lambda}^4 
               +\mathcal{O}\big( {\tilde \lambda}^5\big) \;.
\end{align}
Adding the contributions from $h_0(\tilde\lambda)$ given in
\eqref{eq:sol-h-approx}, and converting $\tilde\lambda$ to $\lambda$
as explained above, we precisely reproduce the term
$(\lambda\,\partial_S\omega\, \partial_T\omega\,\partial_U\omega)^{n-1}$ 
that appears in the expressions for  $\omega^{(n)}$ 
given in  \eqref{eq:higher-orders-3}, \eqref{eq:higher-order-4} (with $a_4 =0$)
and \eqref{eq:omega-5}.

Of course, the functions $\tilde{I}^{(n)}$ also contain non-constant
terms, which contribute to $h_1(u)$. One can verify that they
precisely reproduce all the remaining terms in the expressions
\eqref{eq:higher-orders-3}, \eqref{eq:higher-order-4} (with $a_4 =0$)
and \eqref{eq:omega-5}. This was already pointed out at the end of
section \ref{sec:higher-orders}.

\section{Summary and conclusions}
\label{sec:conclusions}
\setcounter{equation}{0}
In this paper we studied both the holomorphic function that encodes
the Wilsonian effective action of the STU-model of \cite{Sen:1995ff},
as well as its corresponding topological string partition
function. This was done by exploiting the exact duality symmetries of
this model and its invariance under arbitrary permutations of its
three moduli, $S$, $T$, and $U$, which is known as triality. The
topological string partition function can be obtained from the
effective action by carrying out a Legendre transform, as was
explained in \cite{Cardoso:2014kwa}. Based on preliminary calculations
there was the suggestion that the effective action of this STU-model
might be exactly calculable by virtue of its high degree of
symmetry. The fact that the duality transformations act non-linearly
on the function that encodes the Wilsonian action (which itself is not
invariant) was seen as another indication that the answer might even
be unique.  The present paper addresses this and related questions.

A separate motivation for this work was that the connection between
the effective action and the topological string had never been worked
out explicitly for a realistic model.\footnote{%
  Note that we are not implying that this result equals the unique
  expression for the topological string partition function of this
  STU-model, as it is often possible to incorporate additional
  non-holomorphic terms. This possibility was actually discussed in
  \cite{Cardoso:2014kwa}.} %
Especially in the case that exact results can be obtained, one may be
able to obtain valuable insights on this issue. Here we should note
that in this paper we only consider the special-K\"ahler moduli that
describe the vector supermultiplets.

The initial attempt to explicitly determine the Wilsonian action is
described in section \ref{sec:higher-orders}. Because of supersymmetry
the underlying holomorphic function takes the form given in
\eqref{eq:holo-function}, where $\Omega$ is decomposed into an
infinite set of coefficient functions $\omega^{(n)}(S,T,U)$, as is
shown in \eqref{eq:Ups-holo-expansion}.  The exact expression in an
expansion in terms of the gravitational coupling $\lambda=A/(X^0)^2$
is determined by imposing the invariance under dualities and triality
up to $\omega^{(5)}$. While this indicates that an exact determination
is possible, it is also clear that the strategy adopted here could not
easily be continued at arbitrary high orders. Furthermore we should
stress that the results for the holomorphic function encode a complete
$N=2$ supergravity action, which describes all the terms at the
two-derivative level as well as the higher-order couplings to the
square of the Weyl tensor.

At this point it is advantageous to direct the attention to the
topological string partition function, using the results obtained for
the effective action as input. Here one advantage was that the
topological string partition function must be invariant under
duality. Another important is that the moduli for the topological
string transform covariantly and their transformations remain the same
at all orders in the genus expansion, unlike the moduli associated
with the effective action whose transformations keep changing when
proceeding to higher orders.

The results given in \cite{Cardoso:2014kwa} were obtained in
the context of iterative expansions, so that the explicit relation
with the Wilsonian action was only spelled out for low
genus. Here we concentrate on the function $h(\omega;\lambda)$ that
comprises the part of the topological string partition function that
depends holomorphically on the topological string coupling
constant. By following a similar strategy as in the work of
\cite{Alim:2015qma}, we determine an infinite series of terms by
imposing the holomorphic anomaly equation, which can be summed. The
resulting function is denoted by $h_0(u)$, which depends on an
effective duality invariant parameter $u$ that is defined on a Riemann
surface $\mathbb{C}$. The explicit form of this function is given in
equation \eqref{eq:h0-u}. Unfortunately, it turns out that this result
captures only part of the topological string partition function. The
reason is that there exist certain holomorphic invariants expressed in
terms of ordinary holomorphic derivatives of the modular form
$\omega$, which are present in the topological string partition
function. These invariants, denoted by $I^{(n)}$, were already
introduced at the end of section \ref{sec:STU-model-dualities}, and
they are functions of either one of the moduli $S$, $T$, or $U$.

There thus exists an infinite variety of invariants consisting of
arbitrary products of $I^{(m)}(S)$, $I^{(n)}(T)$ and $I^{(p)}(U)$,
which are multiplied by corresponding coefficient functions that
depend on $u$.  In principle this does not imply that these functions
cannot be determined exactly. In fact we have determined a number of
them in section~\ref{sec:addit-contr-h1}. Specifically, we have
determined all terms that are linear and quadratic in the invariants
$I^{(n)}$, as well as some of the cubic terms, and we have verified
their correctness by comparing their holomorphic contributions to
corresponding terms in the effective action.  All these invariants
have been assigned to a second function that we denoted by $h_1(u)$,
so that
\begin{equation}
  \label{eq:h-to-h0-h1-new}
    h(\omega;\lambda) = 
  \omega(S) + \omega(T) + \omega(U)  + h_0(u) 
  + h_1 (u) \;,
 \end{equation} 
 where $h_0(u)$ was defined in \eqref{eq:h0-u} and various
 contributions to $h_1(u)$ have been determined in section
 \ref{sec:addit-contr-h1}.  While we have presented convincing evidence
 that also the function $h_1(u)$ can be determined exactly, the
 situation regarding $h_1(u)$ remains somewhat unsatisfactory, because
 we are dealing with an infinite set of coefficient functions.

 An interesting observation concerns the behaviour of the coefficient
 functions in $h_1(u)$ as compared to $h_0(u)$. The latter function
 has a logarithmic branch cut starting at $u=-1$ and it is vanishing
 at $u=1$, as is shown in equation \eqref{eq:h0-u}. The coefficient
 functions in $h_1(u)$ also exhibit higher-order poles at $u=-1$ and
 zeroes at $u=1$, but in addition they also have poles at $u=0$.  So
 far we have not been able to give an explanation for the presence of
 these new poles.  We note that in the limit where we take the real
 part of two of the three moduli $S, T, U$ large, all these
 singularities disappear.  In this interesting limit, the form of
 $\Omega$ greatly simplifies, as we briefly describe in appendix
 \ref{sec:Large-moduli-limits}.

\subsection*{Acknowledgements}
We would like to thank Cristina C\^amara, Justin David, Thomas Grimm,
Abhiram Kidambi, Suresh Nampuri, Ricardo Schiappa, Ashoke Sen and
Stefan Theisen for valuable discussions. We thank the International
Center for Theoretical Sciences (ICTS, Bangalore), the Mainz Institute
for Theoretical Physics (MITP), the Kavli Institute for the Physics
and Mathematics of the Universe (IPMU), the International Center for
Theoretical Physics (ICTP, Trieste), the Max Planck Institute for
Gravitational Physics (AEI, Golm), and the Kavli Institute for
Theoretical Physics (KITP) for hospitality during various stages of
this work. This research was supported in part by the National Science
Foundation under Grant No. NSF PHY-1748958. This work was partially
supported by FCT/Portugal through UID/MAT/04459/2019, 
and by UGC, Govt. of India, through the special assistance programme F-530/13/DRS/2013-2018 (SAP-I). 

\begin{appendix}
\section{The explicit expressions for $N_{IJ}$ and $N^{IJ}$}
\label{sec:formulae}
\setcounter{equation}{0}
We compute the matrices $N_{IJ}$ and $N^{IJ}$ for the STU-model based on \eqref{eq:class-F}. To this end, we display various components of $F_{IJ}$,
\begin{equation}
  \label{eq:F-IJ}
  F_{00}  = 2\mathrm{i}\, STU\,,\qquad
  F_{01} = -TU\,,\qquad 
  F_{12} = -\mathrm{i}\, U\,.
\end{equation}
Then, using $N_{IJ} = 2\, {\rm Im} F_{IJ}$, we obtain
\begin{equation}
  \label{eq:N-IJ}
  N_{IJ} = \begin{pmatrix}
               2(STU +\bar S \bar T\bar U) & 
               \mathrm{i} (TU-\bar T\bar U) & 
               \mathrm{i} (US -\bar U \bar S) &
               \mathrm{i} ST-\bar S\bar  T)\\[2mm]
               \mathrm{i} (TU-\bar T\bar U) &0    
               &- (U+\bar U) & -(T+\bar T)  \\[2mm]
               \mathrm{i} (US -\bar U\bar S) &  - (U+\bar U)   & 0 
               & -(S+\bar S)  \\[2mm]
               \mathrm{i} (TS-\bar T \bar S)  & - (T+\bar T)
               &  -(S+\bar S) & 0
               \end{pmatrix} \,.
\end{equation}
The inverse matrix $N^{IJ}$ reads,
\begin{equation}  \label{eq:N-IJ-inverse}
  N^{IJ} = \mathrm{e}^\mathcal{K} \begin{pmatrix}
               2   & \mathrm{i} (S-\bar S)  & \mathrm{i} (T-\bar T)
                &\mathrm{i} (U-\bar U)              \\[2mm]
               \mathrm{i} (S- \bar S) & 2\,S\bar S   
               &  - (ST +\bar S\bar T)  & - (US +\bar U\bar S)  \\[2mm]
               \mathrm{i} (T-\bar T) &     - (ST +\bar S\bar T)    & 2\,
               T\bar T 
               &  -  (TU+\bar T\bar U)  \\[2mm]
               \mathrm{i} (U -\bar U)  & - (US +\bar U\bar S) 
               &  - (TU+\bar T\bar U)  & 2\, U\bar U 
               \end{pmatrix} \,,
\end{equation}
where $\det N_{IJ} =- \mathrm{e}^{-2\mathcal{K}}$ and
$\mathrm{e}^{-\mathcal{K}} = (S+\bar S)(T+\bar T)(U+\bar U) = -\bar
X^I N_{IJ} X^J \,\vert X^0\vert^{-2}$.

\section{The structure of $\Omega$ in the large-$T$ and 
large-$U$ limit}
\label{sec:Large-moduli-limits}
\setcounter{equation}{0}

In sections \ref{sec:partial-det-h} and \ref{sec:addit-contr-h1} we
analyzed the structure of the function $h(\omega;\lambda)$ that
describes part of the topological string partition function. As shown
in \eqref{eq:h-to-h0-h1-new} $h(\omega;\lambda)$ is written in terms
of the modular form $\omega$ and two functions $h_0(u)$ and
$h_1(u)$. The function $h_0(u)$, given in \eqref{eq:h0-u},
depends on an effective, duality
invariant coupling constant $u$ defined in terms of the topological
string coupling constant \eqref{eq:u-lambdatilde}, and the function $h_1(u)$ is
decomposed in terms of products of the invariants ${\tilde I}^{(n)}$
with $u$-dependent coefficients that are subject to
\eqref{eq:anom-eq-h1-u}.

In this appendix we want to study $h(\omega;\lambda)$ in the limit
where the real part of two of the three moduli $S, T, U$, say
${\rm Re} \, T$ and ${\rm Re} \, U$, are taken to be large.  For
${\rm Re}\, T \gg 1, {\rm Re}\,U \gg 1$, the derivatives
$\partial_T \omega (T)$ and $\partial_U \omega (U)$ remain finite,
while higher derivatives of $\omega(T)$ and of $\omega(U)$ are
exponentially suppressed.  As follows directly from
\eqref{eq:def-I}-\eqref{eq:serre-derivative}, this leads to
\begin{equation}
  \label{eq:I-appr}
  I^{(n)}(T) \approx \frac{(n-1)!}{2\,\gamma^{n-1}}\,
  \Big(\frac{\partial\omega}{\partial T}\Big)^n \,. 
\end{equation}
Furthermore the (multiple) non-holomorphic derivatives of $\omega(T)$
satisfy,
\begin{equation}
  \label{eq:I-appt-2}
  D_T\omega(T)\approx \partial_T\omega(T)\,, \qquad D_T{\!}^n \omega(T)
  \approx 0\,, \quad (n > 1)\,,
\end{equation}
as follows from \eqref{eq:D-omega}-\eqref{eq:D-Sima}. As a result the
quantities $\tilde I^{(n)}$ defined in \eqref{eq:def-tilde-I}, satisfy
\begin{equation}
  \label{eq:app-3}
   {\tilde I}^{(n)} (T) \approx \frac{(n-1)!}{2\,\gamma^{n-1}}\,.
\end{equation}
Note that there are identical results for $U$, but not
for $S$. 

To determine the function $h(\omega;\lambda)$, we need the functions
$h_0(u)$ and $h_1(u)$. While $h_0(u)$ is explicitly known
(c.f. \eqref{eq:h0-u}), the function $h_1(u)$ must satisfy the
non-linear differential equation \eqref{eq:anom-eq-h1-u}, which in the
above large-moduli limit simplifies dramatically,
\begin{equation}
     \label{eq:odeh1}
   \frac{(u^2 -1)}{2u} \,  \frac{\partial h_1}{\partial u}  - \sum_{n=2}
  \,n\,\tilde{I}^{(n)}(S) \,\frac{\partial
    h_1}{\partial\tilde{I}^{(n)}(S) } 
    = \tfrac12 \gamma \, (u^2 -1) - 2 \gamma \, \frac{u-1}{u+1} \;. 
\end{equation}
Since we are interested in $h_0(u)+h_1(u)$, we consider a linear
differential equation for the sum,
\begin{equation}
  \label{eq:diff-h01}
\frac{\partial(h_0 +h_1)}{\partial u}
 - \frac{2\,u}{u^2-1}  \,   \sum_{n=2}
  \,n\,\tilde{I}^{(n)}(S) \,\frac{\partial
    h_1}{\partial\tilde{I}^{(n)}(S) } 
  = \gamma \, u\,,
\end{equation}   
where we used that $h_0$ does not depend on
$\tilde{I}^{(n)}(S)$. Obviously $h_0+h_1$ can be expanded in powers of
$\tilde{I}^{(n)}(S)$ and by comparing equal powers on the left- and on
the right-hand side of the above equation we derive the $u$-dependence
of the coefficient functions, which involve integration
constants. 
An additional integration constant is determined by using that 
$h_0 + h_1$ vanishes at $u=1$. The final answer for $h_0+h_1$ then takes the form,
\begin{align}
   \label{eq:sol-h01} 
  h_0(u)+h_1(u) =&\, 
  \tfrac12 \gamma \,  \left( u^2 -1 \right)
            + \sum_{m= 2} \alpha_{m}  \, (u^2 -1)^{m} \,
                 \tilde{I}^{(m)}(S)  \nonumber\\[1mm]
        &\, + \sum_{m, n=2} \alpha_{m,n}  \, (u^2 -1)^{m + n}
                 \, \tilde{I}^{(m)}(S)  \, \tilde{I}^{(n)}(S)  \nonumber \\[1mm]
  &\, + \sum_{m, n,p=2} \alpha_{m,n,p}  \, (u^2 -1)^{m + n+p}
           \, \tilde{I}^{(m)}(S)  \, \tilde{I}^{(n)}(S)\,
           \tilde{I}^{(p)}(S) +\cdots \;, 
\end{align}
where the coefficients $\alpha_{m,\ldots}$ are integration
constants. Obviously there is a substantial simplification in the
$u$-dependence of \eqref{eq:sol-h01} as compared to the general terms
that we found in section \ref{sec:addit-contr-h1}. Although in the
latter case all contributions vanish at the perturbative point
$u=1$, the dependence on $u$ in section \ref{sec:addit-contr-h1} is never solely expressed in powers of $u^2-1$.
This behaviour is really the result of the large-moduli limit,
which restricts all the $\tilde{I}^{(p)}(T)$ and $\tilde{I}^{(p)}(U)$
to constants. This induces an enormous rearrangement of terms that
lead to \eqref{eq:sol-h01}. As we will see,
we may compare the results by matching the coefficients of an expansion in
powers of $u-1$.

To illustrate this, let us consider the expressions derived in section
\ref{sec:addit-contr-h1} and analyze how they contribute in the
large-moduli limit to terms linear and quadratic in the
$\tilde{I}^{(n)}(S)$.  {From} \eqref{eq:sol-h01} we know that the results
can be expressed in an expansion in powers of $u-1$, where the lowest
power is known. Likewise the coefficient functions that we determined
in section \ref{sec:addit-contr-h1} can also be expressed as a power
series in $u-1$ and the lowest power should therefore coincide with
the one indicated in \eqref{eq:sol-h01}. We have calculated the
following terms in section \ref{sec:addit-contr-h1}, 
\begin{align}
  \label{eq:summary-6}
  & \sum_{m=2} c_m(u)\, \Big[ \tilde{I}^{(m)}(S) + \cdots\Big] \nonumber\\
  +\,& \sum_{m,n=2} d_{m,n}(u)\, \Big[ \tilde{I}^{(m)}(S)\, \tilde{I}^{(n)}(S) +
    \cdots\Big] \nonumber\\
  +\, & \sum_{m,n=2} e_{m,n}(u) \, \Big[ \tilde{I}^{(m)}(S)\, \tilde{I}^{(n)}(T) +
    \cdots\Big] \nonumber\\
   +\,&p(u) \Big[ \big(\tilde{I}^{(2)}(S)\big)^2\, \tilde{I}^{(2)}(T) +\cdots
      \Big]     \nonumber\\[2mm] 
  +\,&v_{2,2,2}(u) \Big[\tilde{I}^{(2)}(S)\, \tilde{I}^{(2)}(T)\,
    \tilde{I}^{(2)}(U) \,\Big] \,,
\end{align}
where the ellipses denote the extra terms that follow from triality,
while the $u$-dependence of the coefficient functions is known. Note
that the above terms only represent a subset of all possible terms. 

Let us first consider the terms linear in $\tilde{I}^{(n)}(S)$ that
arise in the large-moduli limit of \eqref{eq:summary-6},
\begin{equation}
  \label{eq:alpha-n}
   \Big[c_n(u) + \sum_{m=2} e_{n,m}(u)\, \frac{(m-1)!}{\gamma^{m-1}}
    \Big]\,   \tilde{I}^{(n)}(S)
   + \frac1{4\,\gamma^2} \, \Big[ 2\,p(u) + v_{2,2,2}(u) \Big] \,
    \tilde{I}^{(2)}(S) \,. 
\end{equation}
By comparing powers of $u-1$ it follows that the lowest power of the
above expression must be equal to $(u-1)^n$. However, the only term
that is proportional to $(u-1)^n$ in the above expression is the first
one equal to $c_n(u)$. Assuming that the above equation indeed
contains all the relevant terms, it follows that $\alpha_n$ is
directly related to the value of $(u-1)^{-n} c_n(u)$ at the
perturbative point $u=1$.  This yields the result,
\begin{equation}
  \label{eq:alpha-n}
  \alpha_n = \frac{\gamma^n}{2^n n!}\;. 
\end{equation}

Based on \eqref{eq:summary-6} one can also analyze the terms
of second order in $\tilde{I}^{(m)}(S)$ and compare them to
$\alpha_{m,n}$. Here the large-moduli limit yields, 
\begin{equation}
  \label{eq:tildeI-2-square}
  \Big[d_{2,2}(u) + \frac1{\gamma}\, p(u) \Big] 
  \big(\tilde{I}^{(2)}(S)\big)^2 = \frac{(u-1)^4}{8}\,
  \big(\tilde{I}^{(2)}(S)\big)^2 +\mathcal{O}\big((u-1)^5\big)\,. 
\end{equation}
where we made use of \eqref{eq:d22} and \eqref{eq:pu4}. Upon comparing
this with the corresponding term in \eqref{eq:sol-h01} we derive
\begin{equation}
  \label{eq:1alpha-2-2}
  \alpha_{2,2} = \frac{\gamma^3}{128}\,,
\end{equation}
provided that all relevant terms were included in
\eqref{eq:summary-6}. Using \eqref{eq:summary-6}, one contribution to
the values of the $\alpha_{m,n}$ for $m>2$ can be read off directly from the functions
$d_{m,n}(u)$ given in \eqref{eq:de-(mn)}. However, there may be further contributions to
these values from terms that we have not computed, such as generalizations of the term
proportional to $p(u)$ in \eqref{eq:summary-6}
that involve $\tilde{I}^{(m)}(S)$ with $m>2$.

{From} the result \eqref{eq:sol-h01} one can directly obtain the
large-moduli limit of the function $\Omega$ that encodes the effective
Wilsonian action by invoking \eqref{eq:hole-limit-h}.  From
\eqref{eq:u-lambdatilde} we have
\begin{equation}
  \label{eq:u-lambda} 
  u^2-1  =  \frac{2 \lambda}{\gamma^2} \, D_S \omega \,
  D_T \omega \, D_U \omega \approx \frac{2 \lambda}{\gamma^2} \,
  \partial_S  \omega \, \partial_T \omega \, \partial_U \omega\,,
\end{equation}
where the second equation is the result of the large-moduli limit and
we suppress non-holomorphic corrections. Upon including the terms
linear in $\omega(S)+\omega(T)+\omega(U)$ we then obtain the following
result, 
\begin{align}
   \label{eq:sol-h-omega-lambda} 
  h(\omega;\lambda) =&\, \omega(S)+\omega(T)+\omega(U) 
            +   \frac\lambda{\gamma} \, 
               \partial_S\omega \,\partial_T\omega\,\partial_U\omega
                       \nonumber\\
           &\, + \sum_{m= 2} \alpha_{m}  \,
                \Big(\frac{2\lambda}{\gamma^2}\Big)^m
             (\partial_T\omega\,\partial_U\omega)^{m} \,
                        {I}^{(n)}(S)  \nonumber\\[1mm]
                     &\, + \sum_{m, n=2} \alpha_{m,n}  \, \Big(
                       \frac{2\lambda}{\gamma^2}\Big)^{m+n} 
                    (\partial_T\omega\,\partial_U\omega)^{m + n}
                       \, {I}^{(m)}(S)  \, {I}^{(n)}(S)  \nonumber \\[1mm]
             &\, + \sum_{m, n,p=2} \alpha_{m,n,p}  \,
              \Big(\frac{2\lambda}{\gamma^2}\Big)^{m+n+p} 
             (\partial_T\omega\,\partial_U\omega)^{m + n+p}
                       \, {I}^{(m)}(S)  \, {I}^{(n)}(S)\,
                       {I}^{(p)}(S) +\cdots \;,
\end{align}
which is holomorphic. According to
\eqref{eq:hole-limit-h},
the above expression yields an expansion in
terms of $\lambda$ with coefficients that can be be compared to the
functions $\omega^{(n)}(S,T,U)$ that appear in the function $\Omega$ as
defined in \eqref{eq:Ups-holo-expansion}. We can now use the known
values for $\alpha_2$, $\alpha_3$, $\alpha_4$ and $\alpha_{2,2}$ to
obtain the large-moduli limit of $\omega^{(n)}(S,T,U)$ with $n\leq 5$
and compare the result with the explicit results obtained in section
\ref{sec:higher-orders}. The result takes the form 
\begin{align}
   \label{eq:large-moduli} 
  h(\omega;\lambda) =&\, \omega(S)+\omega(T)+\omega(U) 
                     + \frac\lambda{\gamma} \, 
               \partial_S\omega \,\partial_T\omega\,\partial_U\omega
                       \nonumber\\ 
               &\,  + \frac{\lambda^2}{2\gamma^2}\,
                   \left( \partial_T\omega\,\partial_U\omega\right)^2  \,I^{(2)}(S)   
          + \frac{\lambda^3}{6\gamma^3}\,
                  \left(  \partial_T\omega\,\partial_U\omega\right)^3
                       \,I^{(3)}(S)   \nonumber\\
    &\,  + \frac{\lambda^4}{24\gamma^4}\,
                    \left(\partial_T\omega\,\partial_U\omega\right)^4
                       \, \Big[I^{(4)}(S) + \frac3{\gamma}
      \big[I^{(2)}(S)\big]^2\Big]  +\mathcal{O}(\lambda^5)\,. 
\end{align}
All these terms agree with the results that follow from the large-moduli limit
applied to the corresponding functions $\omega^{(n)}(S,T,U)$ that were
calculated in section \ref{sec:higher-orders}.

\end{appendix}

\providecommand{\href}[2]{#2}


\begin{thebibliography}{10}
%
 \bibitem{Sen:1995ff} 
   A. Sen and C. Vafa, {\it Dual pairs of type II string compactification}, 
                  Nucl. Phys. {\bf B455} (1995) 165--187,
    {\tt hep-th/9508064}.
%
   \bibitem{Gregori:1999ns} A. Gregori, C. Kounnas and P.M. Petropoulos,
     {\it Non-perturbative triality in heterotic and type II N = 2
       strings}, 
     Nucl. Phys. {\bf B553} (1999)  108--132, {\tt hep-th/9901117}.
%
\bibitem{Grimm:2007tm}
  T.W.~Grimm, A.~Klemm, M.~Mari\~no and M.~Weiss,
  {\it Direct integration of the topological string},
  JHEP {\bf 0708} (2007) 058
  [hep-th/0702187 [hep-th]].
%
 \bibitem{deWit:1984pk}
  B. de Wit and A. Van Proeyen, {\it Potentials and
    symmetries of general gauged N=2 supergravity-Yang-Mills
    theory}, Nucl. Phys. {\bf B245} (1984) 89.
%
\bibitem{Cecotti:1988qn} 
  S.~Cecotti, S.~Ferrara and L.~Girardello,
  {\it Geometry of type-II Superstrings and the moduli of
  superconformal field theories},
  Int.\ J.\ Mod.\ Phys.\ A {\bf 4}, 2475 (1989).
%
\bibitem{Cardoso:2008fr}
  G.L.~Cardoso, B.~de Wit and S.~Mahapatra,
  {\it Subleading and non-holomorphic corrections to N=2 BPS black
    hole entropy}, 
  JHEP {\bf 0902} (2009) 006
  [arXiv:0808.2627 [hep-th]].
%
\bibitem{Cardoso:2010gc}
  G.L.~Cardoso, B.~de Wit and S.~Mahapatra,
  {\it BPS black holes, the Hesse potential, and the topological
    string},
  JHEP {\bf 1006} (2010) 052
   [arXiv:1003.1970 [hep-th]].
%
\bibitem{Cardoso:2012nh}
  G.L.~Cardoso, B.~de Wit and S.~Mahapatra,
  {\it Non-holomorphic deformations of special geometry and their applications},
  Springer Proc.\ Phys.\  {\bf 144} (2013) 1
  [arXiv:1206.0577 [hep-th]].
%

\bibitem{Cardoso:2014kwa}
  G.L.~Cardoso, B.~de Wit and S.~Mahapatra,
  {\it Deformations of special geometry: in search of the topological string},
  JHEP {\bf 1409} (2014) 096
  [arXiv:1406.5478 [hep-th]].
%
\bibitem{Bershadsky:1993ta}
  M.~Bershadsky, S.~Cecotti, H.~Ooguri and C.~Vafa,
  {\it Holomorphic anomalies in topological field theories},
  Nucl.\ Phys.\  B {\bf 405} (1993) 279
  [hep-th/9302103].
%
\bibitem{Bershadsky:1993cx} M. Bershadsky, S. Cecotti, H. Ooguri and
                  C. Vafa, {\it Kodaira-Spencer theory of gravity and
                  exact results for quantum string amplitudes},
                  Commun. Math. Phys. {\bf 165} (1994) 311 [hep-th/9309140].
%
\bibitem{Cardoso:2015qhq}
  G.L.~Cardoso and T.~Mohaupt,
  {\it Hessian geometry and the holomorphic anomaly},
  JHEP {\bf 1602} (2016) 161
  [arXiv:1511.06658 [hep-th]].
%
\bibitem{Alim:2015qma} M.~Alim, S.~T.~Yau and J.~Zhou, {\it Airy
    equation for the topological string partition function in a
    scaling limit}, Lett.\ Math.\ Phys.\ {\bf 106} (2016) 719 
  [arXiv:1506.01375 [hep-th]].
%
\bibitem{deWit:2010za} 
  B.~de Wit, S.~Katmadas and M.~van Zalk,
  {\it New supersymmetric higher-derivative couplings: Full N=2
    superspace does not count!} ,
  JHEP {\bf 1101}, 007 (2011)
  [arXiv:1010.2150 [hep-th]].
%
\bibitem{Butter:2013lta} 
  D.~Butter, B.~de Wit, S.~M.~Kuzenko and I.~Lodato,
  {\it New higher-derivative invariants in N=2 supergravity and the
    Gauss-Bonnet term} ,
  JHEP {\bf 1312}, 062 (2013) 
  [arXiv:1307.6546 [hep-th]].
%
\bibitem{Dabholkar:2005dt}
  A.~Dabholkar, F.~Denef, G.W.~Moore and B.~Pioline,
  {\it Precision counting of small black holes},
  JHEP {\bf 0510} (2005) 096
  [hep-th/0507014].
%
\bibitem{Gopakumar:1998ii}
  R.~Gopakumar and C.~Vafa,
  {\it M Theory and topological strings 1} [hep-th/9809187].
%
\bibitem{Gopakumar:1998jq}
  R.~Gopakumar and C.~Vafa,
  {\it M Theory and topological strings 2}   [hep-th/9812127].
%
\bibitem{LopesCardoso:2000qm}
  G.L.~Cardoso, B.~de Wit, J.~K\"appeli and T.~Mohaupt,
  {\it Stationary BPS solutions in N=2 supergravity with $R^2$ interactions},
  JHEP {\bf 0012} (2000) 019,  [hep-th/0009234].
%
\bibitem{Denef:2007vg}
  F.~Denef and G.W.~Moore,
  {\it Split states, entropy enigmas, holes and halos},
  JHEP {\bf 1111} (2011) 129
  [hep-th/0702146].
%
\bibitem{Sen:2011ba}
  A.~Sen, {\it Logarithmic corrections to N=2 black hole entropy: An
    infrared  window into the microstates}, 
  Gen.\ Rel.\ Grav.\  {\bf 44} (2012) no.5,  1207
  [arXiv:1108.3842 [hep-th]].
%
\bibitem{Dedushenko:2014nya}
  M.~Dedushenko and E.~Witten,
  {\it Some details on the Gopakumar-Vafa and Ooguri-Vafa formulas},
  Adv.\ Theor.\ Math.\ Phys.\  {\bf 20} (2016) 1,
  arXiv:1411.7108 [hep-th].
%
\bibitem{Hahn:2015}
   H.~Hahn, {\it Eisenstein series associated with $\Gamma_0(2)$},
   Ramanujan J., {\bf 15}  (2008), 235  
  [arXiv: 1507.04425 [math.NT]].
%
\bibitem{Cardoso:2019avb}
  G.L.~Cardoso, S.~Nampuri and D.~Polini,
  {\it An approach to BPS black hole microstate counting in an $N=2$ STU
  model},   
  arXiv:1903.07586 [hep-th].
%
\bibitem{LopesCardoso:2006ugz} 
  G.L. Cardoso, B.~de Wit, J.~K\"appeli and T.~Mohaupt,
  {\it Black hole partition functions and duality},
  JHEP {\bf 0603}, 074 (2006)
  [hep-th/0601108].
%

%


\end{thebibliography}
\end{document}